\documentclass[trackchanges,twocolumn]{aastex701}

\usepackage{graphicx}
\usepackage{subfig}

\sfcode`A=1000 \sfcode`B=1000 \sfcode`C=1000 \sfcode`D=1000
\sfcode`E=1000 \sfcode`F=1000 \sfcode`G=1000 \sfcode`H=1000
\sfcode`I=1000 \sfcode`J=1000 \sfcode`K=1000 \sfcode`L=1000
\sfcode`M=1000 \sfcode`N=1000 \sfcode`O=1000 \sfcode`P=1000
\sfcode`Q=1000 \sfcode`R=1000 \sfcode`S=1000 \sfcode`T=1000
\sfcode`U=1000 \sfcode`V=1000 \sfcode`W=1000 \sfcode`X=1000
\sfcode`Y=1000 \sfcode`Z=1000

\newcommand{\um}{$\mu$m}
\newcommand{\jwst}{{JWST}}
\newcommand{\vrad}{\ensuremath{v_{\mathrm{rad}}}}
\newcommand{\den}{cm$^{-3}$}
\newcommand{\Ne}{$n_{\mathrm{e}}$}
\newcommand{\Te}{$T_{\mathrm{e}}$}
\newcommand{\vbar}{$v_{\mathrm{Bar}}$}
\newcommand{\varm}{$v_{\mathrm{NArm}}$}
\newcommand{\vblob}{$v_{\mathrm{BC}}$}
\newcommand{\pf}{Pf$\alpha$}
\newcommand{\hii}{\ion{H}{2}}
\newcommand{\kms}{\hbox{km~s$^{-1}$}}

\usepackage{amsmath}
\usepackage{graphicx}
\usepackage{hyperref}

\defcitealias{vonFellenberg2025_extinction}{vF25}
\defcitealias{Fritz2011}{F11}
\newcommand{\Sg}{Sgr~A*}

\begin{document}

\title{Ionization Structure and Metal Enrichment of the Galactic Center Minispiral Observed with JWST}

\author[0000-0001-8921-3624]{Nicole M. Ford}
\affiliation{McGill University, Montreal QC H3A 0G4, Canada}%
\affiliation{Trottier Space Institute, 3550 Rue University, Montréal, Québec, H3A 2A7, Canada}
\email[show]{nicole.ford@mail.mcgill.ca} 

\author[orcid=0000-0001-9641-6550,sname=Balakrishnan, gname=Mayura]{Mayura Balakrishnan}
\affiliation{McGill University, Montreal QC H3A 0G4, Canada}%
\affiliation{Trottier Space Institute, 3550 Rue University, Montréal, Québec, H3A 2A7, Canada}
\email[]{mayura.balakrishnan@mcgill.ca}

\author[orcid=0000-0000-0000-0001,sname=von Fellenberg]{Sebastiano D. von Fellenberg}
\altaffiliation{Feodor Lynen Fellow}
\affiliation{Canadian Institute for Theoretical Astrophysics, University of Toronto, 60 St. George Street, Toronto, ON M5S 3H8, Canada}
\affiliation{Max Planck Insitute for Radioastronomy, auf dem H{\"u}gel 69, Bonn, Germany }
\email{sfellenberg@cita.utoronto.ca}  

\author[0000-0001-6803-2138]{Daryl Haggard}
\affiliation{McGill University, Montreal QC H3A 0G4, Canada}%
\affiliation{Trottier Space Institute, 3550 Rue University, Montréal, Québec, H3A 2A7, Canada}
\email{daryl.haggard@mcgill.ca}

\author[orcid=0000-0003-3503-3446,sname=Michail, gname=Joseph]{Joseph M. Michail}
\altaffiliation{NSF Astronomy \& Astrophysics Postdoctoral Fellow}
\affiliation{Center for Astrophysics | Harvard \& Smithsonian, 60 Garden Street, Cambridge, MA, USA 02138}
\email{joseph.michail@cfa.harvard.edu}

\author[0000-0001-6311-4345]{Yuzhu Cui}
\affiliation{Institute of Astrophysics, Central China Normal University, Wuhan, China}
\email{yuzhu_cui77@163.com}

\author[0000-0002-5599-4650]{Joseph L. Hora}
\affiliation{Center for Astrophysics | Harvard \& Smithsonian, 60 Garden Street, Cambridge, MA, USA 02138}
\email{jhora@cfa.harvard.edu}

\author[0000-0002-8247-786X]{Joey Neilsen}
\affiliation{Department of Physics, Villanova University, 800 Lancaster Avenue, Villanova, PA 19085, USA}
\email{joseph.neilsen@villanova.edu}

\author[0000-0003-0406-7387]{Giacomo Principe}
\affiliation{Dipartimento di Fisica, Universit\'a di Trieste, I-34127 Trieste, Italy}
\affiliation{Istituto Nazionale di Fisica Nucleare, Sezione di Trieste, I-34127 Trieste, Italy}
\affiliation{INAF Istituto di Radioastronomia, Via P. Gobetti, 101, I-40129 Bologna, Italy}
\email{giacomo.principe@inaf.it}

\author[0009-0003-9906-2745]{Tamojeet Roychowdhury}
\affiliation{Department of Astronomy, University of California Berkeley, Berkeley, CA 94704, USA}
\email{tamojeet@iitb.ac.in}

\author[0000-0001-7134-9005]{Nadeen B. Sabha}
\affiliation{Universität Innsbruck, Institut für Astro- und Teilchenphysik, Technikerstr. 25/8, 6020 Innsbruck, Austria}
\email{Nadeen.Sabha@uibk.ac.at}

\author[]{Howard A. Smith}
\affiliation{Center for Astrophysics | Harvard \& Smithsonian, 60 Garden Street, Cambridge, MA, USA 02138}
\email{hsmith@cfa.harvard.edu}

\author[0009-0004-8539-3516]{Zach Sumners}
\affiliation{McGill University, Montreal QC H3A 0G4, Canada}%
\affiliation{Trottier Space Institute, 3550 Rue University, Montréal, Québec, H3A 2A7, Canada}
\email[]{ronald.sumners@mail.mcgill.ca}

\author[0000-0002-9895-5758]{S. P. Willner}
\affiliation{Center for Astrophysics | Harvard \& Smithsonian, 60 Garden Street, Cambridge, MA, USA 02138}
\email{swillner@cfa.harvard.edu}





\begin{abstract}
Sagittarius A* (\Sg) is the nearest quiescent supermassive black hole, and its proximity offers a unique opportunity to study its surrounding fuel supply. We leverage extensive spatial and spectroscopic information provided by the \jwst/MIRI MRS instrument to disentangle mid-infrared ionized gas structures in the central 0.1 parsec of the Galaxy. The Galactic Minispiral's Bar and Northern Arm are revealed by their distinct morphological and kinematic signatures. Several compact ($<1$\arcsec) gas structures including X7 also appear within $\sim 0.05$ parsec of \Sg\ in the plane of the sky, moving with blue-shifted radial velocities $\gtrsim$600~\kms. Fine-structure-line measurements spanning ionization potentials $\sim$7--55~eV are used to constrain the incident radiation field, metal abundances (neon, argon, sulfur, nickel, and iron), and dust depletion/destruction for each identified gas structure. Overall, the Minispiral gas shows mildly super-solar abundances (${\sim}1.8\times$ solar), a Wolf--Rayet star-driven radiation field consistent with ionization parameters spanning $-1.6\lesssim\log U\lesssim-0.6$, and significant nickel and iron dust destruction. In particular, the enhanced ionization state and strong dust processing in the high-velocity compact gas structures suggest that these features may experience localized fast shocks. 

\end{abstract}



\section{Introduction}\label{sec:intro}
\Sg, the supermassive black hole embedded in the Milky Way's Galactic Center (GC) located $\sim$8 kpc from Earth \citep{Ghez2008,GRAVITYCollaboration2019}, is surrounded by a swarm of gas, dust, and stars. It is positioned within the dense nuclear star cluster (NSC) that hosts a large population of hot, UV-emitting Wolf--Rayet (WR) stars, and beyond that, a ring of cold gas called the circumnuclear disk \citep[CND;][]{Vollmer2004,Christopher2005,Hsieh2021}. Interior to the CND lies a collection of warm ionized gas and molecular dust known as the GC ``Minispiral." The Minispiral follows a three-pronged `triskelion' shape composed of the Northern Arm, Eastern Arm, and the Western Arc \citep{Lacy1991,Roberts1993,Zhao2009,Zhao2010}; these filamentary structures act as conduits bridging the CND's cold molecular gas and the hot plasma in the NSC \citep[e.g.,][]{Townes1983,Vollmer2000,Paumard2004,Zhao2010,Cuadra2015,Moussoux2018}. 

Decades of observations with increasing resolution \citep[][and references therein]{Genzel2010,Ciurlo2026} have 
mapped the Minispiral gas primarily in the radio/(sub-)millimeter \citep[e.g.,][]{Balick1974,Wollman1976,Rodriguez1979,Ekers1983,Lo&Claussen1983,YusefZadeh1990,Zhao2009,Zhao2010,Tsuboi2016,Moser2017,Wenner2025} and infrared \citep[IR; e.g.,][]{Lacy1979,Lacy1980,Lacy1982,Roberts1993,Shields1994,Lutz1993,Lutz1996,Stolovy1996,Moultaka2004,Muzic2007,Bhat2022,Dinh2024}. The radio and (sub-)millimeter continua are dominated by thermal bremsstrahlung from ionized gas and by thermal dust emission, respectively. Hydrogen radio recombination lines trace the same ionized gas producing the thermal bremsstrahlung, whereas (sub-)millimeter lines are produced in dense environments where molecular species, such as carbon monoxide, can exist. In contrast, IR line emission includes both recombination and fine-structure lines, with the latter probing the ionization structure and chemistry of the Minispiral gas.

\begin{figure*}[ht!]
    \centering
    \includegraphics[width=\textwidth]{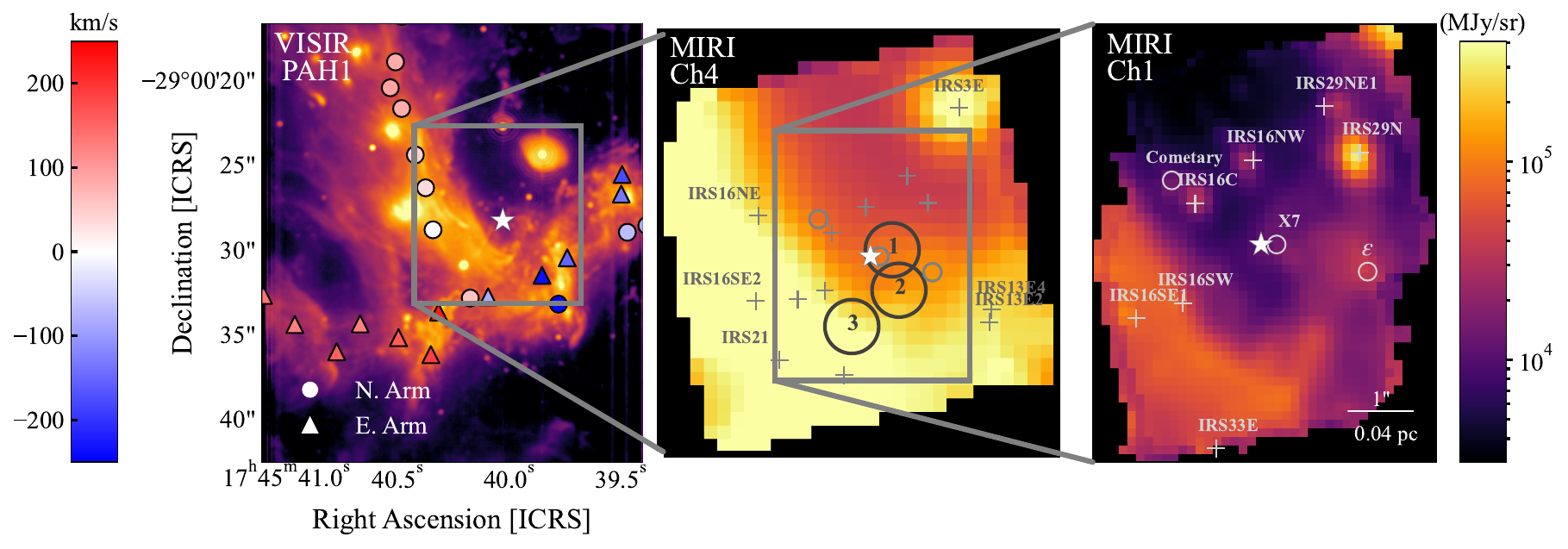}
    \footnotesize 
    \caption{The Galactic Minispiral as seen in the mid-infrared. The panels zoom in from left to right. \textit{Left}: The VISIR PAH1 $8.58$ \um\ image \citep{Dinh2024}, \textit{middle}: MIRI MRS summed flux in spectral channel 4 (Ch4; 17.70--27.90~\um), and \textit{right}: spectral channel 1 (Ch1; 4.9--7.65~\um). The left panel is overlaid with coordinates colored by gas radial velocity derived from Br$\gamma$ \citep{Nitschai2020}. \Sg's position is marked with a white star in all panels. In the middle and right panels, WR star positions \citep{Paumard2006,Ciurlo2023,Dinh2024} are marked with gray crosses, and a selection of compact structures \citep{Dinh2024} are marked with gray circles. Spectral extraction apertures are labeled with black circles.}
    \label{fig:zoomin}
\end{figure*}

The Minispiral's dynamics are governed by the gravitational and radiative effects of \Sg, the NSC, and their interactions. Both the Northern and Eastern Arms trace ionized and kinematically fast gas streamers coming from the CND with velocities of $\sim$100--400~\kms\ along the line of sight, as shown in Figure~\ref{fig:zoomin} \citep{Paumard2004,Zhao2009,Zhao2010,Nitschai2020}. The Northern Arm is inclined towards us with a counter-clockwise rotation \citep[e.g.,][]{Nitschai2020}, and it appears to have two branches \citep{Paumard2004}: an upper branch that wraps closer in to \Sg\ (in projection) and a lower branch that includes the mini-cavity region \citep[e.g.,][]{YusefZadeh1989}. The arms may intersect at an extended interface known as the Bar, which is nearly stationary in radial velocity \citep{Zhao2010}. Based on studies of radio recombination lines and free-free continuum emission---e.g., \cite{Roberts1993, Zhao2010} and the overview by \cite{Ferriere2012}-- the Minispiral gas is thought to have electron densities $n_e \sim 10^4$--$10^5$~\den\ and temperatures $T_e \sim 5\,000$--13\,000~K with the Bar tending to be slightly hotter and denser than the arms. Smaller-scale dusty structures have also been detected, for example the populations of `X' and `G' objects, which can reach radial velocities of several hundred \kms\ \citep[e.g.,][]{Eckart2004,Gillessen2012,Gillessen2026,YusefZadeh2016,Ciurlo2020,Ciurlo2023,Peissker2020,Peissker2023}. 

The ionizing radiation field that bathes the central parsec arises through a combination of past/episodic \Sg\ activity \citep{Koyama1996,Clavel2013,Goldwurm2026}, numerous O/B/WR stars \citep[e.g.,][]{Paumard2004,Martins2007}, and shocks generated by the collisions of orbiting gas parcels and winds \citep{Najarro1997,Moultaka2004,Wang2013}. Using the Infrared Space Observatory (ISO) Short Wavelength Spectrometer (SWS), \cite{Lutz1996} detected ions such as [\ion{Ne}{2}], [\ion{Ne}{3}], [\ion{S}{4}], and [\ion{O}{4}] from the Minispiral gas (see Table~\ref{tab:lines_ip} for line properties). \cite{Paumard2004} traced the Minispiral gas in \ion{He}{1} and Br$\gamma$ emission, finding that it has a roughly uniform composition and an ionization state dependent on proximity to nearby hot stars. GC gas beyond the central parsec remains primarily excited by photoionization, with a secondary contribution from localized shocks \citep[see, e.g.,][]{Simpson2007}.

The Minispiral's chemistry is a product of the star formation histories of both the NSC and the larger-scale nuclear stellar disk \citep[NSD; see, e.g.,][]{NoguerasLara2021,NoguerasLara2023,Schultheis2021,Feldmeier2025}. Abundance measurements indicate that the material feeding the Minispiral has a super-solar gas composition intermixed with dust. IR observations of the broader GC region find mildly super-solar abundances in gas and the majority of stars \citep{Willner1979,Lacy1980,Lester1981,Shields1994,Giveon2002,Do2018}. Constraints from X-ray spectroscopy independently support a metal-enriched environment if solar hydrogen abundance is assumed \citep{Nobukawa2010,Hua2023,Balakrishnan2026}. The elevated $Z$ and, in particular, mildly enhanced $\alpha$ elements \citep[e.g.,][]{Thorsbro2020,Ryde2025,Vermot2025} are consistent with massive star formation and subsequent core-collapse supernovae enriching the GC (\citealt{Pfuhl2014,Feldmeier-Krause2017,Schoedel2020}, but see also \citealt{Chen2023}). 

Both the gas composition and ionizing radiation field influence how efficiently GC gas can cool, clump, and potentially accrete onto \Sg. The ability to form a cold disk \citep{Murchikova2019,Ciurlo2021} in \Sg\ accretion simulations depends strongly on the adopted plasma abundances, illustrating how local chemistry shapes GC dynamical models \citep{Ressler2018,Ressler2020,Cuadra2008,Cuadra2015,Calderon2016,Calderon2025}. 

With the \jwst\ Mid-Infrared Instrument (MIRI)'s Medium Resolution Spectrograph \citep[MRS;][]{Wells2015}, we spatially map numerous fine-structure lines emitted by the Minispiral gas at sub-parsec scales, building on earlier IR studies \citep[e.g.,][]{Lutz1993,Lutz1996,Shields1994,Giveon2002,Paumard2004,Vermot2025}. This allows us to characterize the Minispiral gas' ionization structure and chemical composition, which is important for building an accurate picture of the \Sg\ fuel reservoir.

The paper is structured as follows. Section~\ref{sec:methods} outlines our data reduction process for producing images and spectra from the MIRI MRS cubes. Section~\ref{sec:analysis} describes how we fit mid-IR (MIR) spectral lines and use them to estimate the physical properties of the Minispiral gas. Section~\ref{sec:results} presents measurements of key diagnostic line ratios and element abundances for multiple identified Minispiral gas structures. Section~\ref{sec:disccon} leverages these results to characterize each Minispiral gas structure's ionization state and chemical composition in the context of the GC stars, gas, dust, and \Sg. Finally, Section~\ref{sec:conc} summarizes our key findings.

\section{Data Calibration}\label{sec:methods}
\subsection{Observations and Data Reduction}\label{subsec:datared} 
\begin{figure*}[ht!]
    \centering
    \includegraphics[width=0.985\textwidth]{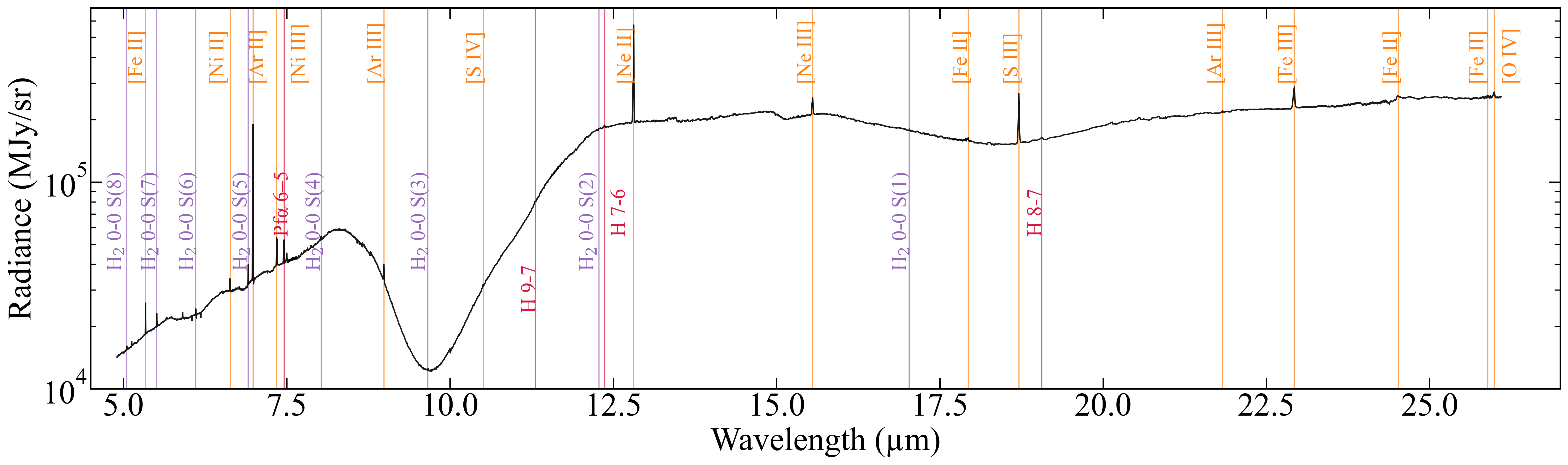}
    \caption{Summed 4.9--27.9 \um\ MIRI MRS observed spectrum extracted from the $\sim$3\arcsec$\times$3\arcsec\ Ch1 footprint, with identified fine-structure (orange), recombination (dark red), and H$_2$ shock emission lines (purple) are labeled.}
    \label{fig:fullspec}
\end{figure*}

As part of \jwst\ Program ID (PID) 4572 \citep{jwst_4572}, we collected three sets of MIRI MRS observations of the GC central $\sim 0.1$ parsec on 2024 April 8-9 and September 6. The dataset consists of two full-spectrum, four-point dither cubes (spectral channels 1--4 with the short, medium, and long grating settings; 22.2 minutes of exposure per setting) and eleven longer exposure observations (spectral channels 1--4 with the short grating; 22.3 hours per channel). The full wavelength ranges covered by channels 1--4 (hereafter abbreviated Ch1--4) are: 4.9--7.65 \um\ (Ch1; spectral resolving power $R\sim3400$), 7.51--11.70 \um\ (Ch2, $R\sim3000$), 11.55--17.98 \um\ (Ch3, $R\sim2300$), and 17.70--27.90 \um\ (Ch4, $R\sim1600$). The full spectrum summed across the Ch1 FoV is shown in Figure~\ref{fig:fullspec}. At \Sg's distance, 1 pc corresponds to roughly 25\arcsec.

We calibrate the data using the JWST pipeline \texttt{jwst} version 1.17 \citep{jwst_pipeline_bushouse2025} with \texttt{pmap} version 1322. In-depth calibration details are provided by \citet[][hereafter \citetalias{vonFellenberg2025_extinction}]{vonFellenberg2025_extinction}, which utilizes the same dataset as this work. In the Level 2 pipeline, we include the residual fringe removal step.\footnote{See \href{https://jwst-pipeline.readthedocs.io/en/stable/jwst/residual_fringe/main.html}{calwebb\_spec2 documentation} for more information.} We run each observation through the Level 3 pipeline separately and incorporate pointing offset corrections to mitigate astrometric errors. Subsequent postprocessing steps include convolution (described below), reprojection to a common WCS, and a flux matching step before the data are averaged over all observations. These steps produce the final cubes we use in this analysis.

We produce two sets of Ch1--4 cubes (eight cubes total): one set maintains the default spatial resolution in each channel, and the other set smooths all channels to the $\sim$0\farcs35 spatial resolution at $17.98$ \um\ (hereafter called the ``native resolution" and ``convolved" data, respectively). For the latter, we apply a 2D Gaussian smoothing kernel assuming the spatial point spread function (PSF) described by the fit from \cite{Law2023}.
Figure~\ref{fig:zoomin} highlights the different fields and spatial resolutions covered in our dataset. In this work, we primarily focus on the spectral properties of the convolved cubes; the native resolution cubes are presented in Appendix~\ref{app:ChMaps}. 

\subsection{Continuum Subtraction}
To isolate line emission, we first model and subtract the continuum. For the native resolution cubes, we subtract on a per-spaxel basis (producing a continuum-subtracted line cube), while for the convolved cubes we use aperture-extracted spectra (Section~\ref{subsec:specex}). 

For each identified emission line, we extract a $\sim 0.15$ \um-wide spectral subset centered on the line's rest wavelength. This is broad enough to encompass both continuum and line emission for the Minispiral's velocity range in our field of view (FoV; see Section~\ref{sec:intro}). To model the continuum, we first mask out the 0.02--0.08~\um-wide ``line region" (equivalently, a velocity range of $\sim$1000~\kms). The exact width depends on the line's broadness and the spectral resolution. In addition to the line mask, we exclude prominent emission/absorption features 
that could bias the continuum fit and any pixels with un-physical flux values (e.g., NaN pixels flagged by the pipeline). 

We define a polynomial model to describe the continuum emission on either side of the masked line region. The polynomial is either linear or quadratic depending on the spectral resolution and complexity of the local continuum shape. The fitted model is interpolated to span the full $\sim$0.15~\um\ width (including the line region) and then subtracted from the spectrum. This results in a continuum-subtracted spectral subset containing only the line emission (see Figure~\ref{fig:specdefringe}). 

Before fitting models to the continuum-subtracted lines, we correct for residual fringing in the spectrum. In addition to the \jwst\ pipeline de-fringing mentioned in Section \ref{subsec:datared}, we apply the de-fringing procedure described in detail by \citetalias{vonFellenberg2025_extinction}. The method combines Factor Analysis and Gaussian Processes to `learn' the fringing profile spatially and spectrally for each line's masked continuum region. The fringe profile is subsequently interpolated and subtracted to isolate the line emission. 

\section{Analysis}\label{sec:analysis}
We detect numerous emission and absorption features; this work focuses on 13 forbidden fine-structure emission lines spanning ionization potentials (IPs) of 7.9--54.93~eV (Table~\ref{tab:lines_ip}). All reported lines reach peak signal-to-noise (S/N) $\gtrsim10$. The lines originate from iron (Fe), nickel (Ni), sulfur (S), argon (Ar), neon (Ne), and oxygen (O). These elements fall in one of two groups: (1) $\alpha$ elements (sulfur, argon, neon, and oxygen), formed by helium nuclei capture in massive stars and during core collapse supernovae, and (2) iron-peak elements (nickel and iron), formed primarily during Type Ia supernovae. We also report four hydrogen recombination lines: H 6--5 (\pf) 7.46\,\um, H 9--7 11.31\,\um, H  (Hu$\alpha$) 7--6 12.37\,\um, and H 8--7 19.06\,\um. These lines are useful for measuring absolute abundances relative to hydrogen (Section~\ref{subsec:abunds}). A more extensive list of detected recombination lines is provided by \citetalias{vonFellenberg2025_extinction}. The detected H$_2$ lines (Figure~\ref{fig:fullspec}) are the subject of a separate study, but preliminary analysis indicates that they trace shocked molecular gas that is primarily not co-spatial with the Minispiral structures considered in this work.

\begin{deluxetable}{crrc}
    \tablecaption{\jwst\ MIRI MRS emission lines used in this work.\label{tab:lines_ip}}
    \tablehead{
    \colhead{Identification} & 
    \colhead{$\lambda_{\mathrm{rest}}$} & \colhead{IP} & \colhead{$\log n_\mathrm{crit}$} \\
    \colhead{} & \colhead{\micron} & \colhead{(eV)} & \colhead{(\den)}
    }
     \startdata
     \lbrack \ion{Fe}{2}\rbrack & 5.34 & 7.90 & 2.47\\
     \lbrack \ion{Ni}{2}\rbrack & 6.64 & 7.64 &  5.90\\
     \lbrack \ion{Ar}{2}\rbrack & 6.99 & 15.80 & 5.57\\
     \lbrack \ion{Ni}{3}\rbrack & 7.35 & 18.20 &  5.96\\
     H 6--5 & 7.46 & 13.60 &  \nodata\\
     \lbrack \ion{Ar}{3}\rbrack & 8.99 & 27.60 &  5.23\\
     \lbrack \ion{S}{4}\rbrack & 10.51 & 34.80 &  4.68\\
     H 9--7 & 11.31 & 13.60 &  \nodata\\
     H 7--6 & 12.37 & 13.60 &  \nodata\\
     \lbrack \ion{Ne}{2}\rbrack & 12.81 & 21.60 &  5.74\\
     \lbrack \ion{Ne}{3}\rbrack & 15.56 & 40.96 &  5.26\\
     \lbrack \ion{Fe}{2}\rbrack & 17.94 & 7.90 & 4.69\\
     \lbrack \ion{S}{3}\rbrack & 18.71 & 23.30 &  4.00\\
     H 8--7 & 19.06 & 13.60 &  \nodata\\
     \lbrack \ion{Fe}{3}\rbrack & 22.93 & 16.19 &  4.54\\
     \lbrack \ion{O}{4}\rbrack & 25.89 & 54.93 &  3.95\\
     \lbrack \ion{Fe}{2}\rbrack & 25.99 & 7.90 &  4.08\\
    \enddata
    \tablecomments{Lines detected with peak S/N $\gtrsim 10$, including rest wavelengths ($\lambda_{\mathrm{rest}}$) and ionization potentials (IPs) from the Atomic Spectra Database \citep[ASD;][]{Kramida2022} of the National Institute of Standards and Technology (NIST; but see also Appendix~\ref{app:FeIII}), and critical densities ($n_\mathrm{crit}$) from \texttt{PyNeb} for a fixed $T_e=7\,500$~K. IP is the minimum energy required to remove the most loosely bound electron(s) and produce the ion. $n_\mathrm{crit}$ is the density at which an ion is equally likely to radiatively decay or collisionally de-excite.}
\end{deluxetable}

\subsection{Spectral Extraction}\label{subsec:specex}
To maximize the line signal, we extract integrated spectra from three circular apertures named Regions 1--3 (Figure~\ref{fig:zoomin}) in the convolved cubes, using \texttt{jdaviz} \citep{jdaviz2026}. The regions are defined to have a radius of 2 pixels ($\sim0\farcs7$), comparable to the spatial PSF Full-Width at Half-Maximum (FWHM) for these cubes' resolution.  

Regions 1--3's positions are chosen to be fully visible in all four MIRI MRS channels and to contain no faulty pixels (e.g., no NaNs or negative fluxes prior to continuum subtraction, though see Appendix~\ref{subsec:cruciform}). For this reason, Region 2 is partially overlapping Region 1. If it were positioned any lower to avoid overlap, it would intersect with pixels that are saturated in the bright [\ion{Ne}{2}] 12.81 \um\ line and masked out. Region 3 exceptionally contains two such masked pixels, but this does not impact our analysis (Appendix~\ref{app:NeII}). Likewise, we position the regions to avoid noisy pixels associated with the MIR bright WR stars IRS 33E, IRS 16SE, and IRS 16SW. 

All regions overlap with portions of the Minispiral's Northern Arm and Bar (Section~\ref{sec:intro}). Region 1 also contains Sgr~A* \citep[based on its J2000.0 radio coordinates $266\fdg4168$, $-29\fdg0078$;][]{Reid2020}.
Region 3 is offset approximately 2\arcsec\ south and 2\arcsec\ east of Region 1 to capture emission from where the Northern Arm intersects (in projection) with the Eastern Arm/Bar. 

\subsection{Line Fitting}\label{subsec:linefits}
\begin{figure}
    \centering
    \includegraphics[width=0.45\textwidth]{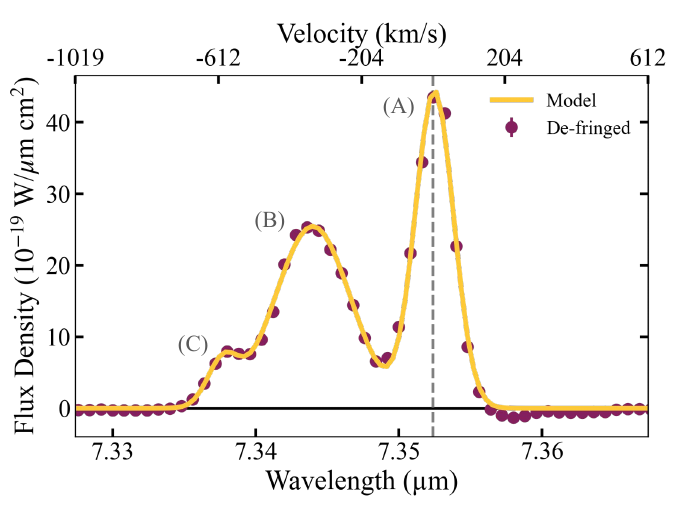}
    \caption{\footnotesize An example de-fringed and continuum-subtracted spectral line ([\ion{Ni}{3}] 7.35 \um) extracted from Region 2 (Figure~\ref{fig:zoomin}). 
    De-fringed data (maroon) is overlaid with the best fit multi-Gaussian model (yellow line) with fit components labeled A--C, and the black line represents the continuum model.}
    \label{fig:specdefringe}
\end{figure}

We construct models of the continuum-subtracted, integrated emission lines that consist of one to three Gaussian components and fit them to the data using \texttt{scipy}'s \texttt{curve\_fit} function. For each line listed in Table~\ref{tab:lines_ip}, we select the model whose reduced $\chi^2$ statistic is closest to 1 and whose components are detected with at least $5\sigma$ confidence. In almost all cases, the spectra are best fit by double or triple Gaussian models. An example model fit for [\ion{S}{4}] 10.51 \um\ is shown in Figure~\ref{fig:specdefringe}.

Ions whose lines are better fit with a single Gaussian include [\ion{Fe}{2}] and [\ion{O}{4}]. For [\ion{Fe}{2}], the lack of additional fit components is likely due to most of the detectable iron emission being further ionized to [\ion{Fe}{3}]. For [\ion{O}{4}] 25.89 \um, residual fringing prohibits a confident detection of more than one fit component (see \citealt{Crouzet2025} for further discussion of fringe modeling difficulties in Ch4). Likewise, we do not detect [\ion{Ne}{5}] 14.32 \um\ or 24.32 \um, which is unsurprising given this ion's large IP (97 eV).

Each fit component is described by a central wavelength ($\lambda_\mathrm{obs}$), width ($\sigma$), and amplitude. From these parameters, we calculate the radial velocity \vrad\ $= c\frac{\lambda_\mathrm{obs} - \lambda_{\mathrm{rest}}}{\lambda_{\mathrm{rest}}}$, where $\lambda_{\mathrm{rest}}$ is the rest wavelength and $c$ is the speed of light. In this \vrad\ definition, the sign convention is positive for redshift and negative for blueshift. We use \texttt{astropy.coordinates} to transform \Sg's sky position to Galactic coordinates and correct \vrad\ from the barycentric frame to the local standard of rest. We also calculate $\mathrm{FWHM}_{\mathrm{model}} \approx 2.355\sigma$ and correct for the MIRI MRS line spread function (LSF) at the corresponding wavelength, using $\mathrm{FWHM}_{\mathrm{line}} = ({\mathrm{FWHM}_{\mathrm{model}}^2 - \mathrm{FWHM}_{\mathrm{LSF}}^2})^{0.5}$ with $\mathrm{FWHM}_{\mathrm{LSF}} = c/R$, where the spectral resolution $R(\lambda_\mathrm{obs}) = 4603 - 128 \times \lambda_\mathrm{obs}$(\um) \citep{Argyriou2023}. For a line to be ``resolved", it must satisfy $\mathrm{FWHM}_{\mathrm{model}} > \mathrm{FWHM}_{\mathrm{LSF}}$. 
If $\mathrm{FWHM}_{\mathrm{model}} \sim \mathrm{FWHM}_{\mathrm{LSF}}$, we assume the line is not resolved and report $\mathrm{FWHM}_{\mathrm{LSF}}$ as an upper limit.

The statistical uncertainty of each fit parameter is estimated as the standard deviation of a Monte Carlo (MC) simulation of the model with added RMS noise. We additionally account for
the \texttt{jwst} pipeline version wavelength solution (FLT-8) uncertainty of up to 10~\kms\ \citep{Argyriou2023} when reporting wavelength-dependent measurements. 

Each best-fit model component is analytically integrated to obtain the line flux. To de-redden the flux, we interpolate the wavelength-dependent absolute extinction ($A_\lambda$) posterior distribution output from \citetalias{vonFellenberg2025_extinction}'s model. We use the posterior median to extinction-correct the flux, and treat the $A_\lambda$ uncertainty as the standard deviation of the full 105 $A_\lambda$ posterior samples. We assume that $A_\lambda$ is constant across each emission line and for all velocity components in each aperture. The defringed fit uncertainty and the extinction correction uncertainty are added in quadrature to obtain the total flux uncertainty, with extinction dominating the error budget by at least an order of magnitude. Final extinction-corrected fluxes and uncertainties are reported in units of W\,cm$^{-2}$\,$\mu$m$^{-1}$ for ease of comparison with literature studies such as \cite{Lutz1996}. The full fit results and de-reddened fluxes for Regions 1--3 are presented in Appendix~\ref{app:linefits}. To demonstrate the effect of the assumed extinction curve, Appendix~\ref{app:ext} provides an alternate de-reddening calculation using the \citet[][hereafter \citetalias{Fritz2011}]{Fritz2011} model.

The spectral-fit components each exhibit a characteristic Doppler shift. The three components considered in this work show only small velocity differences across Regions 1--3, which allows us to identify the components by their averaged $\vrad \sim 0$~\kms, $-270$~\kms, and $-570$~\kms. Based on the Minispiral's known kinematic properties and spatial morphology (Section~\ref{sec:intro}, Figures~\ref{fig:ChMaps},~and~\ref{fig:Region1fits}), we associate each Doppler shifted fit component with a particular gas structure, labeled in Figure~\ref{fig:specdefringe}: (A) the Bar, (B) the Northern Arm, and (C) an assortment of faint, blueshifted structures. 

\begin{figure*}[t!]
    \centering
    \includegraphics[width=0.9\textwidth]{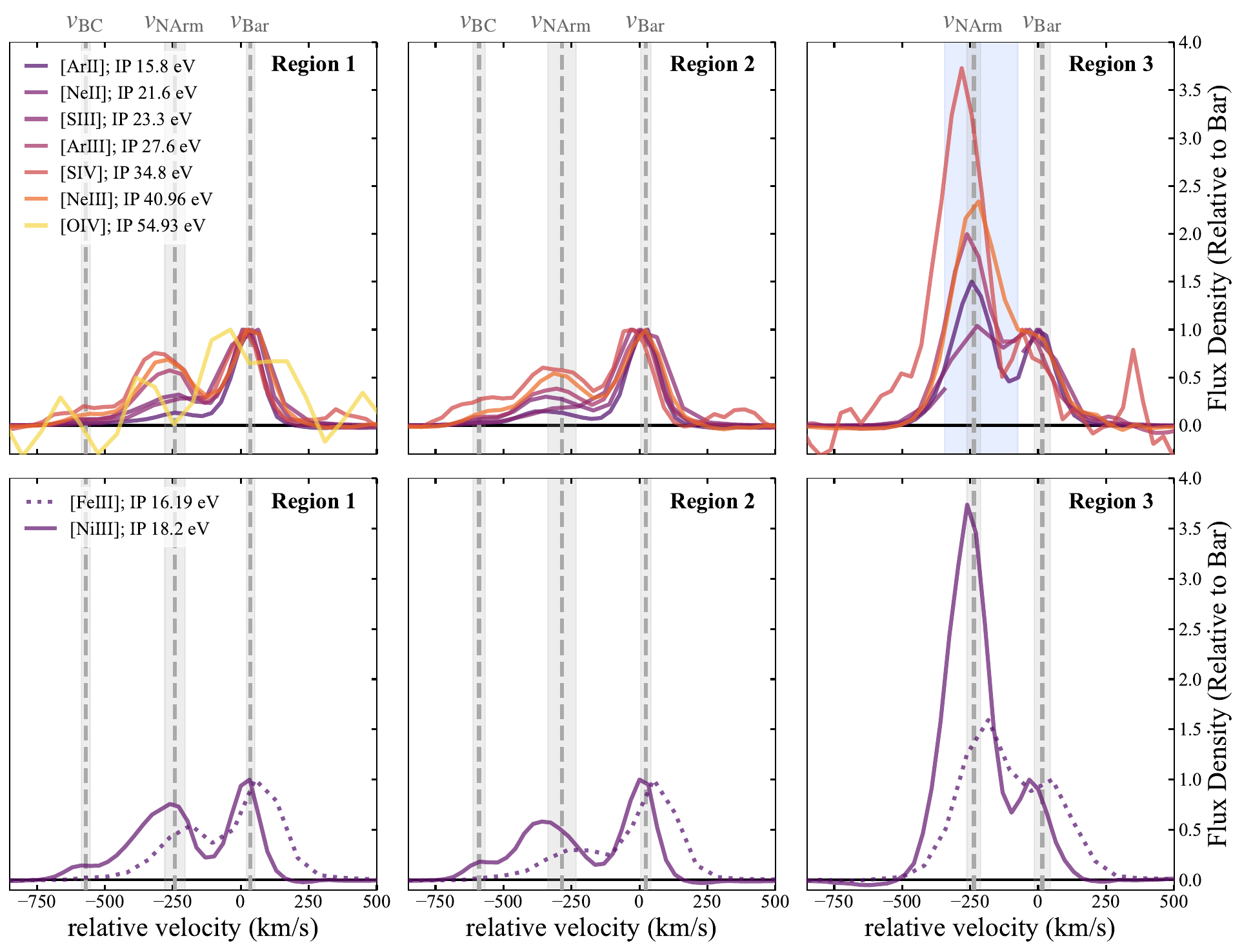}
    \caption{\footnotesize Continuum-subtracted, de-fringed spectra with peak S/N $\gtrsim 10$ extracted from Regions 1--3 (Figure~\ref{fig:zoomin}). Lines are fit by up to three Gaussians with different Doppler shifted \vrad\ (dashed gray lines), corresponding to the Bar (\vbar\ $\sim 0$~\kms), Northern Arm (\varm\ $\sim -270$~\kms), and BC structures (\vblob\ $\sim -570$~\kms). All lines are colored by IP and expressed in relative flux density units, normalized to the Bar's peak flux. \textit{top}: $\alpha$ elements. \textit{bottom}: [\ion{Fe}{3}] 22.93 \um\ and [\ion{Ni}{3}] 7.352 \um. See Appendix~\ref{app:FeIII} for further discussion of rest wavelength uncertainties for these lines. Dashed gray lines indicate each fit component's \vrad\ averaged across lines and $1\sigma$ standard deviation.
    The saturated portion of Region 3's [\ion{Ne}{2}] 12.81 \um\ line is marked with a blue shaded box. The Northern Arm and BC structures have relatively stronger higher IP lines than lower IP lines when compared to the Bar. They also show strongly enhanced nickel flux relative to iron when compared to the Bar.}
    \label{fig:apertures}
\end{figure*}

The blueshifted structures (fit component C) appear only in spectra extracted from Regions~1 and~2 (see Figure~\ref{fig:apertures}). The fitted \vrad\ resembles the predicted orbital velocity for the compact X7 object \citep[$\sim$500--600~\kms;][]{Ciurlo2023}. The tip of X7 inferred from \cite{Ciurlo2023}'s orbital model is marked in Figure~\ref{fig:zoomin}; it falls in Region 1, with a tail that extends across both Regions 1 and 2. Additional compact clumps appear in Regions 1 and 2, exhibiting similarly large blueshifts as X7 (Appendix~\ref{app:ChMaps}). For the remainder of this work, we collectively refer to these as blueshifted compact (BC) structures.

\subsection{Abundance Calculation}\label{subsec:abundmeth}
We convert the extinction-corrected line fluxes to ionic abundances for a chosen \Ne,\Te\ prescription using \texttt{PyNeb} \citep{Luridiana2015, Morisset2020, Mendoza2023}. The requisite atomic data are included in \texttt{PyNeb} and supplemented with [\ion{Ni}{2}] atomic data from \texttt{Chianti} v11.0.2 \citep{DelZanna2021,Dufresne2024}. One version of the abundance calculation uses fixed $n_e = 10^3$~\den\ and $T_e = 7\,500$~K \citepalias{vonFellenberg2025_extinction}, and another version varies \Ne\ and \Te: $n_e = 10^5$~\den, $T_e = 13\,000$~K for the Bar; $n_e = 3 \times 10^4$~\den, $T_e = 6\,000$~K for the Northern Arm; and $n_e = 4 \times 10^4$~\den, $T_e = 7,000$~K for the BC structures. The varying \Ne,\Te\ prescription is more physically motivated based on previous measurements (Section~\ref{sec:intro}), but the fixed \Ne,\Te\ prescription provides a useful metric to assess how the assumed physical conditions affect the abundances. Throughout this work, we assume Case~B recombination with coefficients from \cite{Storey1995}, and the same \Te,\Ne\ prescription for all metal and recombination lines.

To derive absolute ionic abundances for each gas structure in each aperture, the ionic line fluxes must be normalized to a hydrogen recombination line or some other measure of ionization rate (see Table~\ref{tab:lines_ip}). We choose the recombination line whose $A_{\lambda}$ most closely matches that of each ionic line based on the \citetalias{vonFellenberg2025_extinction} as well as the \citetalias{Fritz2011} models (Appendix~\ref{app:ext}), thereby minimizing the differential extinction ($\Delta A_\lambda$) between the two lines. This reduces the normalized abundance's dependence on the assumed extinction model. For example, $A_{\lambda}$ predicted for [\ion{Ne}{3}] differs by almost half a magnitude between the \citetalias{vonFellenberg2025_extinction} and \citetalias{Fritz2011} models. After normalizing to H 7--6, however, the extinction correction largely cancels out for both models (Table~\ref{tab:extinction}). 

Unlike the Bar and Northern Arm, the BC structures are only bright enough to be detected in the H 6--5 recombination line; nevertheless, their other recombination line fluxes can still be inferred. First, we derive $\Delta A_\lambda$ between the desired recombination line and H 6--5 by comparing the observed recombination-line flux ratio with the intrinsic emissivity ratio predicted by \texttt{PyNeb}. This procedure is independent of the assumed absolute extinction model. We perform this calculation for the Bar and Northern Arm and adopt the average $\Delta A_\lambda$ for each aperture. We then use this value to rescale the measured H 6--5 flux of the BC structures and, together with the predicted emissivities, infer the reddened flux of the undetected recombination line. 
Finally, we recover its intrinsic flux by applying the model-based absolute $A_\lambda$ at the desired line's wavelength.

\texttt{PyNeb} follows the historical convention of normalizing by H$\beta$ 0.486~\um. Although H$\beta$ lies far outside MIRI's wavelength range, its flux can be inferred from the chosen de-reddened MIR hydrogen recombination line flux using the corresponding \texttt{PyNeb} emissivity ratio. The H$\beta$ and ionic line fluxes are then combined with our chosen \Ne,\Te\ prescription in \texttt{PyNeb} to calculate the ionic abundance. All relevant ionic abundances are summed to estimate the gas phase element abundance (denoted as $X_i$ for element $i$, hereafter solar-normalized in log form). 

When calculating $X_i$, all gas ionization states must be accounted for.   
If an ionization state(s) is not observed, its ionic abundance must be inferred using an ionization correction factor (ICF). For neon and argon, the IP required to reach the next ionization state is $\gtrsim$~60~eV, high enough that the ICF is likely negligible.\footnote{Additionally, our [\ion{O}{4}] detection is too weak to yield a reliable $X_\mathrm{O}$ estimate.} In contrast, sulfur, iron, and nickel may be more affected. The observed ion emission strengths indicate that unobserved S$^{+}$ ($\rm IP = 10$~eV), Fe$^{3+}$ ($\rm IP = 30$~eV), and Ni$^{3+}$ ($\rm IP = 35$~eV) may contribute substantially to their respective $X_i$. For these elements, we select an ICF range informed by \hii-region models. The measured S$^{2+}$ + S$^{3+}$ ionic abundance needs an ICF of 1.15--1.25, based on a survey of Galactic \hii\ regions \citep{MartinHernandez2002}. The iron ICF varies significantly based on the model, likely due to the chosen atomic data or physical condition assumptions \citep[][and references therein]{MendezDelgado2024}. For the measured Fe$^{+}$ + Fe$^{2+}$ ionic abundance, the likely ICF range is 2.0--3.7 \citep{Lutz1993,Rodriguez2005,Izotov2006,Simpson2007,MendezDelgado2024}. This range is validated by the \cite{Lutz1993} Minispiral mini-cavity study, in which they infer an ICF $=4$. For nickel, an indirect estimate of the ICF is $\sim$1.5--2, based on \cite{DelgadoInglada2016}'s study of \hii\ regions with \Ne,\Te\ comparable to our GC assumptions (Section~\ref{sec:intro}).

Abundance statistical uncertainties are estimated using a MC approach. For each aperture, line (including both the fine-structure and recombination lines), and gas structure, we draw 103 random samples from Gaussian distributions defined by the line flux and uncertainty.\footnote{For the BC structures in recombination lines other than H 6--5, we also MC sample from the inferred $\Delta A_\lambda$ relative to H 6--5 and its standard uncertainty.} Ionic and elemental abundances are computed for each sample using the approach described above. These are converted to log space and normalized to solar abundances by subtracting randomly sampled values from \citet[][drawn in log space from their reported uncertainties, typically $\lesssim 0.05$~dex]{Asplund2021} yielding distributions of $X_i$. We adopt the 16th--84th percentiles of the MC distributions as ${\pm}1\sigma$ uncertainties. To account for the additional uncertainty associated with unobserved ionization states, we also compute the abundance spread after multiplying the observed $X_i$ by the minimum/maximum of the chosen ICF range.

\section{Results}\label{sec:results}
\subsection{Detected Spectral Features}\label{subsec:LineProfs}
The de-fringed, extinction-corrected lines extracted from Regions 1--3 are illustrated in Figure~\ref{fig:apertures}. The relative strengths of the Bar, Northern Arm, and BC structures for a given line depend on both the extraction aperture's spatial position and the line's IP. The Northern Arm and BC structures show higher excitation than the Bar as evidenced by the relative strengths of the higher-IP lines [\ion{S}{4}] and [\ion{Ne}{3}] (see the upper row of Figure~\ref{fig:apertures}, and Section~\ref{subsec:lineratios}). 

The brightest portion of the Northern Arm in our FoV overlaps with Region 3 (Figure~\ref{fig:zoomin}) and shows consistently elevated line fluxes compared to the Bar. In addition to the fine structure lines shown in Figure~\ref{fig:apertures}, \pf\ shows the same trend: the Northern Arm's \pf\ flux is comparable to the Bar's in Regions~1 and~2 but rises almost an order of magnitude above the Bar's in Region 3 (see Table~\ref{app:tab:fullfits}).

[\ion{Ni}{3}] 7.35 \um\ is unusually bright relative to [\ion{Fe}{3}] 22.93 \um\ in the Northern Arm and BC structures compared to the Bar, as shown in the lower row of Figure~\ref{fig:apertures}. Nickel and iron possess the lowest IPs in our dataset (Table~\ref{tab:lines_ip}). The Northern Arm [\ion{Ni}{3}]'s normalized amplitude in Region 3 is so strong that it surpasses all other lines shown in Figure~\ref{fig:apertures} (see also Section~\ref{subsubsec:NiFe}). 

\subsection{Excitation Diagnostics}\label{subsec:lineratios}
\begin{figure}
    \centering
    \includegraphics[width=\linewidth]{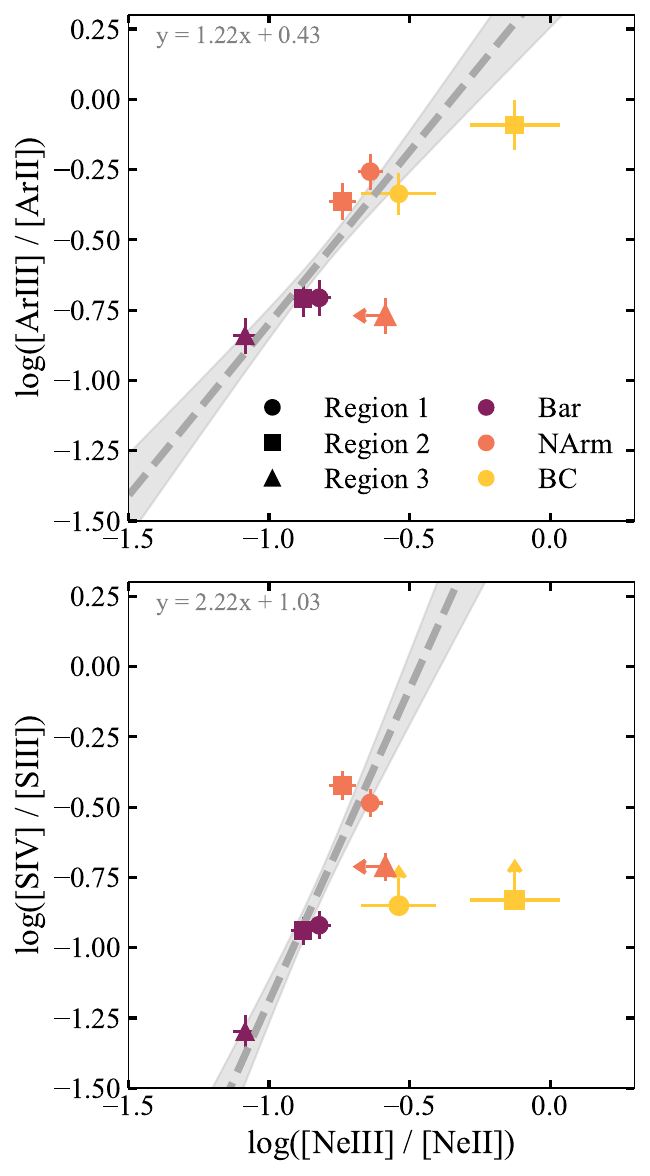}
    \caption{\footnotesize Line ratios comparing different ionization states of argon ([\ion{Ar}{2}] 15.8 eV and [\ion{Ar}{3}] 27.6 eV), sulfur ([\ion{S}{3}] 23.3 eV and [\ion{S}{4}] 34.8 eV), and neon ([\ion{Ne}{2}] 21.6 eV and [\ion{Ne}{3}] 40.96 eV). Ratios are calculated for the Bar (maroon), Northern Arm (orange), and BC structures (yellow) in the spatial apertures labeled Regions 1 (circles), 2 (squares), and 3 (triangles). The line of best fit and $1\sigma$ confidence band are shown in gray. Arrows indicate 3$\sigma$ limits (due to either [\ion{Ne}{2}] saturation or [\ion{S}{3}] non-detection). The BC structures are more ionized than the Northern Arm, which is more ionized than the Bar; Regions 1 and 2 are always more ionized than Region 3.
    }
    \label{fig:ratios}
\end{figure}

Individual emission lines can be combined into flux ratios that trace different ionization states of the gas. Ratios of ions from the same element are particularly useful because they minimize sensitivity to abundance variations (Section~\ref{subsec:abunds}) and instead primarily indicate the hardness and strength of the ionizing radiation field \citep[Section~\ref{subsec:ionization}; see also][]{Genzel1998,Thornley2000,Giveon2002,Verma2003,Snijders2007,Zhang2025}. Three observed $\alpha$ elements (argon, sulfur, and neon) have ion pairings that sample a significant IP range (Table~\ref{tab:lines_ip}). Figure~\ref{fig:ratios} compares the argon and sulfur flux ratios to neon. 

The BC structures are more ionized than the Northern Arm, which is more ionized than the Bar across all line ratios. This is expected from the normalized lines' flux--IP correlation found in Section~\ref{subsec:LineProfs}, and it is further illustrated in Figure~\ref{fig:ratios}, where the Northern Arm and BC structures' flux ratios are above and to the right of the Bar's. Furthermore, the ratios extracted from Regions 1 and 2 are elevated compared to those from Region 3. Not only is the Northern Arm more ionized than the Bar, but there is also a trend of increasing ionization for regions located closer to \Sg\ (in projection; see Figure~\ref{fig:zoomin}). This trend persists for the BC structures, with their Region 2 emission being the most ionized of all the identified gas features.

Using orthogonal distance regression, we fit a linear relation to the de-reddened line ratios from all apertures and gas structures (excluding any unconstrained data points). Though the BC structures in Region 2 exhibit the largest deviation ($\sim-1.6\sigma$, trending towards a smaller slope), it is not statistically significant.

\subsection{Element Abundances}\label{subsec:abunds}
\begin{figure}
    \centering
    \includegraphics[width=\linewidth]{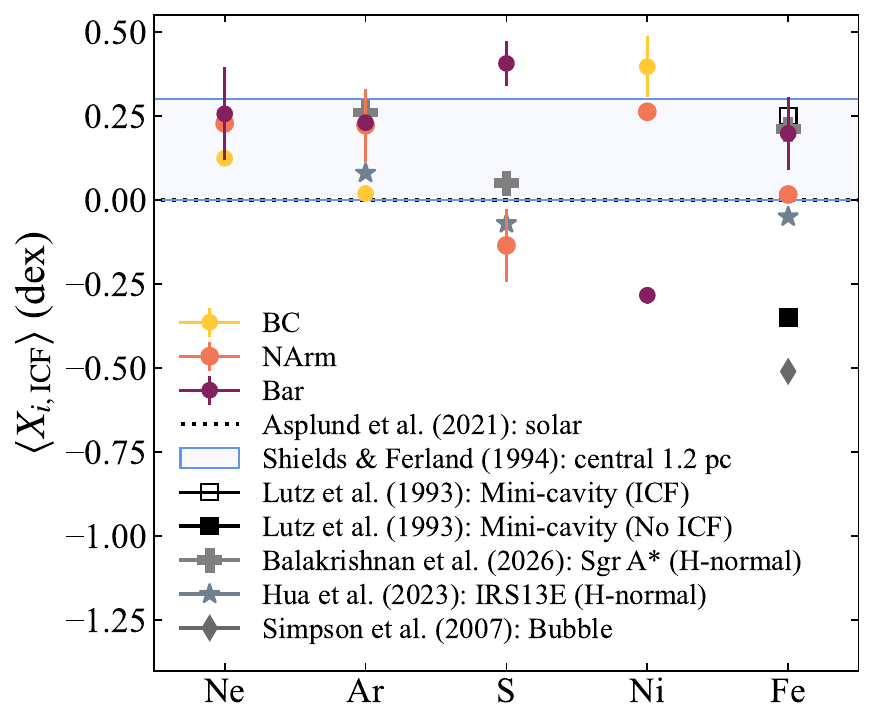}
    \caption{\footnotesize $\langle X_{i\mathrm{,ICF}} \rangle$ for neon, argon, sulfur, nickel, and iron in the varied \Ne,\Te\ prescription. Abundances are given in dex relative to solar. This figure uses the lower limit on the ICF corrections, while the full range of $\langle X_{i} \rangle$ with and without ICF corrections is provided in Table~\ref{tab:abundancesavg}. Uncertainties are given as the standard deviation across Regions 1--3.
    GC literature values are included for the central 1.2 parsec ionized gas (\citealt{Shields1994}; blue bar), the mini-cavity (\citealt{Lutz1993} with and without their inferred ICF; empty and filled black squares, respectively), the Bubble (\citealt{Simpson2007}; gray diamond), the IRS 13 X-ray plasma (\citealt{Hua2023}; gray stars), and the central parsec X-ray plasma (\citealt{Balakrishnan2026}; gray cross). We find mildly super-solar argon and neon, sub- to super-solar sulfur and nickel, and mildly super-solar gas-phase iron.}
    \label{fig:AvgAbund}
\end{figure}

We calculate $X_i$ using two \Ne,\Te\ prescriptions (Section~\ref{subsec:abundmeth}) for the following elements: neon (using [\ion{Ne}{2}] and [\ion{Ne}{3}]), sulfur (using [\ion{S}{3}] and [\ion{S}{4}]), argon (using [\ion{Ar}{2}] and [\ion{Ar}{3}]), iron (using [\ion{Fe}{2}]\footnote{We choose [\ion{Fe}{2}] 17.94 \um\ because it is the least likely to be contaminated by low temperature photodissociation region emission along the line of sight \citep{Kaufman2006,Simpson2007}.} and [\ion{Fe}{3}]), and nickel (using [\ion{Ni}{2}] and [\ion{Ni}{3}]). One version of $X_i$ is calculated using only detected ions ($X_\mathrm{i,noICF}$) and another version is corrected with the ICF schemes described in Section~\ref{subsec:abundmeth} ($X_\mathrm{i,ICF}$). $X_i$ is broadly consistent across Regions 1--3, and Table~\ref{tab:abundancesavg} provides the average $\langle X_i \rangle$ and standard deviation for each element. Individual $X_i$ for all regions, ICFs, and \Ne,\Te\ prescriptions are reported in Table~\ref{app1:tab:abundances} for cases where the listed ionic lines are detected (see also Section~\ref{subsec:gascomp} and Appendix~\ref{subsec:NeTe}). For the remainder of the discussion, we focus on the varied \Ne,\Te\ prescription and apply the lower ICF limits as a conservative estimate of the total gas phase $X_i$. Figure~\ref{fig:AvgAbund} illustrates where each element's $\langle X_{i\mathrm{,ICF}} \rangle$ falls in relation to 0 (corresponding to $\langle X_{i\mathrm{,ICF}} \rangle = X_\sun$ in log space).

In comparison to $X_\sun$, the Minispiral gas in the central $\sim$0.1 parsec is mildly enhanced in argon and neon, with a total gas phase $\langle X_{\mathrm{Ar,ICF}} \rangle \sim \langle X_{\mathrm{Ne,ICF}} \rangle \sim 0.26$ ($1.8 \times$ solar). The spread is up to $\sim$0.1~dex ($\sim$1.4--2.2$\times$ solar) depending on the chosen element and aperture (Table~\ref{tab:abundancesavg}). 
The BC structures tend towards slightly lower $\langle X_{i\mathrm{,ICF}} \rangle$ than the Bar or Northern Arm, but that may simply be due to their strong ionization producing an increased quantity of undetected [\ion{Ne}{4}] and/or [\ion{Ar}{4}] (Section~\ref{subsec:abundmeth}). 
Overall, $\langle X_{\mathrm{Ar,ICF}} \rangle$ and $\langle X_{\mathrm{Ne,ICF}} \rangle$ are consistent with MIR observations of both the central parsec \citep{Lacy1980,Shields1994,Giveon2002} and further out in the GC (e.g., the Arched filaments from \citealt{Simpson2007} and the Radio Arc from \citealt{RodriguezFernandez2005}). 

The refractory elements (sulfur, nickel, and iron) range from mildly sub- to super-solar. Sulfur and nickel can be enhanced or depleted in different gas structures (Section~\ref{subsubsec:NiFe}). Finally, iron shows comparable enhancement as argon and neon, with a spread of $\sim$0.3~dex depending on both the aperture and the gas structure. As seen in Figure~\ref{fig:AvgAbund}, $\langle X_{\mathrm{Fe,ICF}} \rangle$ falls precisely within the measured range observed for the minicavity  \citep{Lutz1993}, IRS 13E \citep{Hua2023}, and the stellar wind plasma around \Sg\ \citep{Balakrishnan2026}.

\begin{deluxetable*}{lccccc}
\tablecaption{Measured element abundances averaged across Regions 1--3.\label{tab:abundancesavg}}
\tablehead{
\colhead{Model ($n_e$ [\den]; $T_e$ [K])} &
\colhead{Element} & 
\colhead{ICF} &
\colhead{$\langle X_\mathrm{Bar} \rangle$} & 
\colhead{$\langle X_\mathrm{NArm} \rangle$} & 
\colhead{$\langle X_\mathrm{BC} \rangle$}
}
\startdata
$10^3$; 7\,500 & Ne & No & $0.425 \pm 0.141$ & $0.107 \pm 0.032$ & $0.054 \pm 0.009$ \\
             & Ar & No & $0.383 \pm 0.017$ & $0.073 \pm 0.111$ & $-0.068 \pm 0.024$ \\
             & S  & No & $-0.477 \pm 0.068$ & $-0.987 \pm 0.108$ & \nodata \\
             & S  & Yes & $[-0.416,-0.380]$ & $[-0.927,-0.890]$ & \nodata \\
             & Ni & No & $-0.231 \pm 0.029$ & $-0.028 \pm 0.025$ & $0.181 \pm 0.091$ \\
             & Ni & Yes & $[-0.055,0.070]$ & $[0.148,0.273]$ & $[0.357,0.482]$ \\
             & Fe & No & $-0.559 \pm 0.103$ & $-0.776$ & \nodata \\
             & Fe & Yes & $[-0.258,0.009]$ & $[-0.475,-0.208]$ & \nodata \\
\tableline
$10^5$; 13\,000 (Bar) & Ne & No & $0.256 \pm 0.138$ & $0.228 \pm 0.031$ & $0.133 \pm 0.023$ \\
$3\times10^4$; 6\,000 (Northern Arm) & Ar & No & $0.230 \pm 0.013$ & $0.222 \pm 0.109$ & $0.028 \pm 0.031$ \\
$4\times10^4$; 7\,000 (BC) & S & No & $0.345 \pm 0.067$ & $-0.196 \pm 0.110$ & \nodata \\
                         & S & Yes & $[0.406,0.442]$ & $[-0.135,-0.099]$ & \nodata \\
                         & Ni & No & $-0.459 \pm 0.024$ & $0.086 \pm 0.023$ & $0.231 \pm 0.097$ \\
                         & Ni & Yes & $[-0.283,-0.158]$ & $[0.262,0.387]$ & $[0.396,0.521]$ \\
                         & Fe & No & $-0.103 \pm 0.109$ & $-0.285$ & \nodata \\
                         & Fe & Yes & $[0.198,0.465]$ & $[0.016,0.283]$ & \nodata \\
\enddata

\tablecomments{Two different \Ne,\Te\ prescriptions are used (Section~\ref{subsec:abunds}). $\langle X_\mathrm{i,noICF} \rangle$ is calculated from the measured ionic abundances only, while $\langle X_\mathrm{i,ICF} \rangle$ is a range computed by multiplying $\langle X_\mathrm{i,noICF} \rangle$ by the lower and upper limits of the adopted ICF range (see Section~\ref{subsec:abundmeth}). 
For the Northern Arm, [\ion{Fe}{2}] is only detected in Region 3. We consider $\langle X_\mathrm{i,ICF} \rangle$ with the ICF lower limits to be a conservative estimate of the total gas-phase abundances.
}
\end{deluxetable*}

\section{Discussion}\label{sec:disccon}
\subsection{Gas Composition}\label{subsec:gascomp}
Based on our inferred gas phase abundances for the volatile $\alpha$ elements neon and argon, we adopt a total average Minispiral metal abundance of $\langle X_\mathrm{i,total} \rangle \sim 0.26$ ($1.8\times$ solar). Although there are empirical relations that scale these $\alpha$ elements to infer the oxygen abundance $X_\mathrm{O,total}$ \citep[a proxy for $Z$;][]{FernandezOntiveros2021,Rogers2026}, future studies are needed to extend these mid-infrared relations from classical \hii\ regions to the GC environment (see also Section~\ref{subsec:metal}).

Below, we consider $X_i$ variations for the different Minispiral structures' refractory elements and compare our findings to existing GC stellar abundance measurements.


\subsubsection{Refractory Elements}\label{subsubsec:NiFe}
$\langle X_{\mathrm{S,ICF}} \rangle$ may be sub- to super-solar depending on the gas structure. Because sulfur is more refractory than argon and neon, it could experience increased dust depletion that leads to an inferred smaller gas phase $\langle X_{\mathrm{S,ICF}} \rangle$. It also has a low $n_\mathrm{crit}$ (Table~\ref{tab:lines_ip}), which makes $\langle X_{\mathrm{S,ICF}} \rangle$ especially sensitive to the assumed \Ne\ (Section~\ref{subsec:NeTe}). A similar explanation applies to $\langle X_{\mathrm{Fe,ICF}} \rangle$, albeit with a smaller scatter between the Bar and Northern Arm. Despite potential biases due to dust depletion and \Ne, $\langle X_\mathrm{S,ICF} \rangle \sim \langle X_\mathrm{Fe,ICF} \rangle \sim 0.11$--0.15 when averaged across the Bar and Northern Arm (Table~\ref{tab:abundancesavg}), in agreement with the findings of \cite{Giveon2002,Simpson2007,RodriguezFernandez2005}. 

There is a relative enhancement of nickel in the Northern Arm and BC structures compared to the Bar. The main suspects for distorting $\langle X_\mathrm{Ni,ICF} \rangle$ are the atomic data and assumed physical conditions. Line fluorescence could bias [\ion{Ni}{2}] emission, but it is unlikely to impact the [\ion{Ni}{3}] ion that dominates $\langle X_\mathrm{Ni,ICF} \rangle$. Incorrect atomic data could specifically bias the inference of Ni$^{2+}$'s abundance \citep{MendezDelgado2021}. However,
the de-reddened [\ion{Ni}{3}] flux---normalized by the inferred H$\beta$ flux (Section~\ref{subsec:abundmeth})---is almost double in the Northern Arm compared to the Bar (Table~\ref{app:tab:fullfits}) prior to imposing atomic data assumptions. 
Nickel should also not be skewed by our choice of \Ne\ so long as $n_e < n_\mathrm{crit}\sim$~10$^6$~\den\ (Section~\ref{subsec:abundmeth} and Table~\ref{tab:lines_ip}). Although the nickel ICF is uncertain (Section~\ref{subsec:abundmeth}), the Bar would require $\rm ICF \gtrsim 3.5$ to bring its $X_{\rm Ni,ICF}$ into agreement with the Northern Arm's; given that the Bar appears to be the least ionized (Section~\ref{subsec:lineratios}), such a large ICF is difficult to justify. Overall, we conclude that nickel should not be biased in a way that produces the observed $\langle X_\mathrm{Ni,ICF} \rangle$, and so the relative enhancement is most likely due to different intrinsic gas-phase abundances.

Shocks are very effective at freeing refractory elements from dust, as seen for iron in the Northern Arm's mini-cavity \citep{Lutz1993} and the Galactic Bubble \citep{Simpson2007}, both with inferred $v_\mathrm{shock} \sim 50$--100~\kms. Table~\ref{tab:PhysProp} gives estimated dust-depletion factors ($\delta X_i$) for nickel and iron.
Typically, only 5--10\% of iron exists in the gas phase in dusty GC environments, corresponding to $\delta X_\mathrm{Fe} \sim-1.5$ \citep{Simpson2007}. By comparison, $-0.24\lesssim \delta X_\mathrm{Fe}\lesssim-0.06$ for both the Bar and Northern Arm. The Bar has $-0.54\lesssim\delta X_\mathrm{Ni} < \delta X_\mathrm{Fe}$, while the Northern Arm and BC structures show essentially no nickel depletion. These depletion factors indicate that dust has been strongly processed throughout the central $\sim0.1$~pc of the Minispiral, with refractory elements being significantly less depleted than in typical GC environments. The comparatively higher $\langle X_\mathrm{Ni,ICF} \rangle$ in the Northern Arm and BC structures is consistent with more extensive shock-driven dust destruction than in the Bar.

\subsubsection{Comparison to GC Stellar Abundances}
GC stellar population observations indicate that the majority of NSC stars have $Z \sim 1$--2~$Z_\sun$ \citep{Feldmeier-Krause2017,Feldmeier-Krause2020,Do2015,Do2018,Anastasopoulou2023}. 
For the central parsec WR stars' natal gas, \cite{Martins2007} infer an initial $Z \sim 1$--2~$Z_\sun$. These values agree with our abundance measurements for the Minispiral when taking appropriate ICFs into account (Table~\ref{tab:abundancesavg} and Section~\ref{subsubsec:NiFe}). 

The central parsec WR stellar winds' composition also appears mildly super-solar, under the assumption that light elements (hydrogen, carbon, nitrogen) have solar abundances. \cite{Hua2023,Hua2025} and \cite{Balakrishnan2026} measure $X_\mathrm{Ar} \sim X_\mathrm{S} \sim X_\mathrm{Fe} \sim 0$--$0.3$ (1--2$\times$ solar; see Figure~\ref{fig:AvgAbund} gray stars and crosses, respectively). The shared composition of the NSC/WR sources and the observed Minispiral gas suggests these objects originate from similar material.

A caveat to the inferred WR stellar wind X-ray abundances is their sensitivity to the assumed light-element composition. \cite{Hua2023,Hua2025} propose an alternate scenario motivated by the assumption that WR stars burn through their hydrogen supply and synthesize carbon/nitrogen (e.g., \citealt{Crowther2007} and references therein). In this hydrogen-depleted case, they find that the WR winds have mildly sub-solar $Z$, even though the stars initially formed from super-solar $Z$ gas.
Our measurements are thus in close agreement with the hydrogen-normal scenario, and in some tension with the hydrogen-depleted hypothesis. 

\subsection{Gas Ionization}\label{subsec:ionization}
Line flux ratios that probe different IP thresholds can constrain the hardness and shape of the underlying ionizing spectral energy distribution (SED) as well as the ionization parameter \citep[$U$; e.g.,][]{Genzel1998,Giveon2002,Zhang2025}. $U$ is defined as 
\begin{equation}
    U = \frac{Q_{\rm H}}{4\pi r^2n_{\rm H}c}
\label{eq:U}\end{equation}
where $Q_{\rm H}$ is the total number of H ionizing photons, $r$ is the distance between the emitting gas and the ionizing source, and $n_{\rm H}$ is the hydrogen density (here taken as $n_{\rm H}\approx n_e$). 

Below, we compare our measured neon ratios (Figure~\ref{fig:ratios} and Table~\ref{app:tab:fullfits}) with predictions from a set of \cite{Martins2007}'s CLOUDY photoionization simulations \citep{Gunasekera2025}, where the simulated SED is the combined radiation field from $\sim30$ WR stars in the central parsec. We initially adopt the same assumptions as \cite{Martins2007}, namely that $Q_{\rm H} = 6 \times 10^{50}$ s$^{-1}$ and $r = 0.5$ pc. This leaves \Ne\ as the remaining free parameter, which is used to solve for $U$.

We consider three \Ne\ and $U$ regimes with predicted neon ratios from \cite{Martins2007}'s simulations:
\begin{enumerate}
    \item Small $n_e$ ($3\times10^3$ \den) and large $\log U~ (-0.6)$:\\
    $\log \lbrack$\ion{Ne}{3}]/[\ion{Ne}{2}] $\sim 0.1$
    \item Medium $n_e$ ($10^4$~\den) and Medium $\log U$ (-1.2):\\
    $\log \lbrack$\ion{Ne}{3}]/[\ion{Ne}{2}] $\sim -0.3$
    \item Large $n_e$ ($10^5$~\den) and Small $\log U$ (-1.6):\\ $\log \lbrack$\ion{Ne}{3}]/[\ion{Ne}{2}] $\sim -1$
\end{enumerate} 

The Bar aligns with scenario (3): we observe $\log$[\ion{Ne}{3}]/[\ion{Ne}{2}] $\sim [-1,-0.8]$ and assume $n_e = 10^5$~\den\ (in the varied \Ne,\Te\ prescription; Section~\ref{subsec:abundmeth}). The Northern Arm instead falls between scenarios (2) and (3), with $-1.6 < \log U < -1.2$: we measure $\log$[\ion{Ne}{3}]/[\ion{Ne}{2}] $\sim [-0.7,-0.6]$, and the assumed $n_e = 3\times10^4$~\den. The Northern Arm's inferred larger $U$ is supported by its enhanced argon and sulfur ratios compared to the Bar's (Figure~\ref{fig:ratios}). Finally, the BC structures fall between scenarios (1) and (2): $\log$[\ion{Ne}{3}]/[\ion{Ne}{2}] $\sim [-0.5,0]$. However, our assumed \Ne\ is larger by a factor of $\sim$5--10 than in either scenario. We use the \Ne\ value inferred from X7's Br$\gamma$ emission in \citealt{Ciurlo2023}, but the BC structures may be less dense than we assumed. 

Alternatively, a modification to one of the other parameters in Equation~\ref{eq:U} could explain the observed neon ratio. For example, \cite{Martins2007} find that neon is primarily ionized by one WR star called IRS 16SE2 (Figure~\ref{fig:zoomin}). The chosen $\sim$1\arcsec\ width of Regions 1--3 samples gas along the inner rim of the Northern Arm that should be closest to IRS 16 SE2 \citep[see Figure~\ref{fig:zoomin};][]{YusefZadeh1993,Paumard2004}. If we relax our initial assumptions, the Bar's lower neon ratios (and inferred smaller $U$) could also be caused by it being farther away (increasing $r$) and/or more shielded from IRS 16SE2 (effectively decreasing $Q_{\rm H}$). The reverse of this (closer proximity/less shielding) could explain the BC structures' inferred larger $U$.

The ionizing radiation field may receive additional contributions from other source(s) besides the WR stars. Based on work by \cite{Balakrishnan2026b}, additional ionizing radiation is unlikely to originate from \Sg's accretion flow in its quiescent or flaring state. Their \texttt{CLOUDY} simulations predict that the peak MIR line fluxes arising from Sgr~A*'s circumnuclear gas are too low to be detected by \jwst\ MIRI. Unlike the accretion flow, fast radiative shocks are known to generate energetic photons capable of exciting MIR lines \citep[e.g., ][]{Allen2008,Zhang2025,Zhang2026}. Given the BC structures' $\vrad\sim 500$--600~\kms, they are likely to experience substantial ram pressure capable of generating these types of shocks.

Beyond comparisons to simulations, we can also compare our line ratios with previous central parsec observations. Our neon ratios are consistently greater than those from studies with larger FoVs. In their ISO observations, \cite{Lutz1996, Lutz1999} find $\log$[\ion{Ne}{3}]/[\ion{Ne}{2}]~$=-1.3$, whereas we observe $-1\lesssim\log$~[\ion{Ne}{3}]/[\ion{Ne}{2}]~$\lesssim$~$0$ in all cases (Figure~\ref{fig:ratios}). The MIRI field used in this work (3.2\arcsec--6.6\arcsec\ $\times$ 3.7\arcsec--7.7\arcsec\ depending on the spectral channel) is contained within ISO's larger aperture ($\sim$14\arcsec$\times$~20\arcsec). Similarly, \cite{Vermot2025} measured $\log$[\ion{Ne}{3}]/[\ion{Ne}{2}] $\sim -0.9$ for a spatially integrated aperture comparable to our full MIRI FoV. For studies that report fluxes from a much larger aperture, the sampled gas may experience varied levels of shielding and ionizing source separations that can lower the overall average neon ratio.

\subsection{Interpreting the Minispiral Gas}
The Minispiral Bar, Northern Arm, and BC structures have different inferred ionization states but similarly enriched metal abundances, as summarized in Table~\ref{tab:PhysProp}. The Minispiral gas being mildly super-solar suggests that it shares a common origin with the majority of the surrounding GC material (Section~\ref{subsec:gascomp}), including the NSC and young massive WR stars. This is consistent with an in-situ star-formation scenario \citep[e.g.,][]{Bonnell2008,HobbsNayakshin2009} in which a dense gas disk forms around \Sg\ and subsequently collapses to produce the observed stellar population. Although similar abundances could arise if the stars and gas formed farther out in the GC and migrated inward together, such scenarios are generally disfavored for the young massive stellar population \citep[e.g.,][]{Zwart2003,Paumard2006,Feldmeier-Krause2015,vonFellenberg2022}. 

Our measured metal abundances in the central $\lesssim$ 0.1 parsec can inform composition assumptions in future gas-dynamics and accretion simulations \citep[e.g.,][]{Cuadra2015,Calderon2016,Ressler2020,Calderon2025}. While the $\alpha$ elements have abundances $\sim$1.8$\times$ solar, the gas-phase iron-peak elements are sensitive to dust and ICF assumptions. Dust depletion and destruction may play an important role in gas clumps' ability to cool and fall onto \Sg\ \citep[e.g.,][]{Omukai2005,Ciotti2007,Namekata2014}. \cite{Calderon2025} find that the $Z \approx 3~Z_{\sun}$ regime may represent a threshold that regulates whether or not gas can cool efficiently into a \Sg\ accretion disk. Our inferred abundances are below this threshold. Future simulations that take into account our gas-phase abundances and dust depletion may help elucidate the gas cooling efficiency.

\begin{deluxetable}{lccc}
\tablecaption{Physical properties for the Bar, Northern Arm, and BC structures.\label{tab:PhysProp}}
\tablehead{
\colhead{} & \colhead{Bar} & \colhead{Northern Arm} & \colhead{BC}
}
\startdata
$n_e$ (\den) & $10^5$ & $3\times10^4$ & $4\times10^4$ \\
$\log U$ & $-1.6$ & $[-1.6,-1.2]$ & [$-1.2,-0.6$] \\
$\langle X_{i,\mathrm{tot}} \rangle$  & $\sim0.26$ & $\sim0.26$ & $\sim0.26$ \\
$\delta X_\mathrm{Ni}$  & $-0.54$ & 0 & 0 \\
$\delta X_\mathrm{Fe}$ & $-0.06$ & $-0.24$ & \nodata \\
\enddata
\tablecomments{Values are based on \Ne,\Te\ from the varied \Ne,\Te\ prescription. The ionization parameter ($\log U$) is inferred from the neon line ratio  and expressed as a range when the measured ratio falls between different theoretical $\log U$ regimes (Section~\ref{subsec:ionization}).  Metal abundances $\langle X_{i,\mathrm{total}} \rangle$ are in log form and solar normalized.
The dust depletion factors $\delta X_\mathrm{Fe}$ and $\delta X_\mathrm{Ni}$ are defined as $\delta X_i\equiv$ $X_\mathrm{i,gas} - X_\mathrm{i,total}$ (in log space, assuming $X_\mathrm{i,total}=0.26$). $\delta X_i=0$ indicates that the element is fully in the gas phase, while $\delta X_i<0$ implies depletion into dust grains. When the gas-phase $X_\mathrm{i,ICF}$ exceeds the inferred $X_\mathrm{i,total}$, we set $\delta X_i$ to zero. The values shown are for the respective ICF lower limits (Section~\ref{subsec:abundmeth}).}.
\end{deluxetable}

\section{Conclusion}\label{sec:conc}
This study constrains the composition and ionization state of three Minispiral gas features distinguished by radial velocity: the Bar, Northern Arm, and fast-moving gas parcels that we call blueshifted compact (BC) structures.

At sub-parsec scales in the Minispiral, we infer metal abundances of $\sim 1.8\times$ solar. Variations between elements are attributed to the assumed physical conditions and/or dust depletion. 
The weak depletion of gas-phase nickel and iron indicates that the Minispiral dust has been strongly processed, consistent with efficient dust destruction by shocks. Gas-phase nickel is particularly enhanced in the Northern Arm and BC structures compared to the Bar, which may be due to more extensive shock-driven dust destruction in those regions.

The ionizing radiation field is generally explainable by the combined flux of the central WR stars \citep{Martins2007}. Variations in this field as a function of aperture position and gas structure can be attributed to a combination of different gas densities, distances or shielding from ionizing sources, and shocks or other external radiation sources.

The BC structures are particularly intriguing because small gas clumps may be the main source of fuel for \Sg\ accretion \citep{Witzel2014,Gillessen2012,Calderon2025,Gillessen2026}. The comparable metal abundances between the BC structures and the surrounding Minispiral gas suggest a common origin. Enhanced ionization and dust destruction may also indicate that localized fast shocks are influencing the BC structures. Future higher S/N measurements of IR line emission at IP $\gtrsim 41$ eV (e.g., [\ion{O}{4}] or [\ion{Ne}{5}]) would help diagnose the roles of photoionization versus shocks in these small scale gas features.


\begin{acknowledgments}
A special acknowledgment is made in dedication to Dr. Giovanni Fazio, a valued colleague whose prolific career has helped shape this work and the fields of high energy and infrared astrophysics. The authors also thank Rainer Schödel for providing the VISIR image used in Figure~\ref{fig:zoomin}, which was originally published by \cite{Dinh2024}. 

NMF, DH, MB, ZS acknowledge support from the Canadian Space Agency (23JWGO2A01 and 25JWGO4A01), the Natural Sciences and Engineering Research Council of Canada (NSERC) Discovery Grant program, the Canada Research Chairs program, the Fondes de Recherche Nature et Technologies (FRQNT) Centre de recherche en astrophysique du Québec, and the Trottier Space Institute at McGill. NMF acknowledges funding from FRQNT and NSERC doctoral scholarships. ZS is supported by the Chalk-Rowles fellowship. MB acknowledges support from NSERC's Banting Postdoctoral Fellowship Program (CIHR AWARD BPF 200617-267964). SDvF gratefully acknowledges the support of the Alexander von Humboldt Foundation through a Feodor Lynen Fellowship and thanks CITA for their hospitality and collaboration. SDvF is supported by NSERC, the Canadian Space Agency (23JWGO2A01), and by a grant from the Simons Foundation (MP-SCMPS-00001470). JM is supported by an NSF Astronomy and Astrophysics Postdoctoral Fellowship under award AST-2401752. TR acknowledges funding support from the Deutscher Akademischer Austauschdienst (DAAD) Working Internships in Science and Engineering (WISE) program.

This research was supported by the International Space Science Institute (ISSI) in Bern, through ISSI International Team project \#24-610, and we thank the ISSI team for their generous hospitality. This work was supported by grant \url{https://doi.org/10.69777/377645} from the Fonds de recherche du Québec and by the Center for Research in Astrophysics of Québec (AstroQuébec). This work is based on observations made with the NASA/ESA/CSA James Webb Space Telescope. These observations are associated with program \#4572. The observations are available at MAST (\dataset[doi:10.17909/cnmm-gv89
]{\doi{10.17909/cnmm-gv89}}). Support for program \#4572 was provided by NASA through a grant from the Space Telescope Science Institute, which is operated by the Association of Universities for Research in Astronomy, Inc., under NASA contract NAS 5-03127.

\end{acknowledgments}




%
\facilities{JWST}

\software{astropy \citep{astropy:2013,astropy:2018,astropy:2022}, scipy \citep{Virtanen2020}, jwst \citep{jwst_pipeline_bushouse2025}}, jdaviz \citep{jdaviz2026}


\appendix
\restartappendixnumbering
\section{Channel Maps}\label{app:ChMaps}
\begin{figure*}
    \centering
    \includegraphics[trim=0mm 4mm 0mm 0mm, clip=true, width=0.75\textwidth]{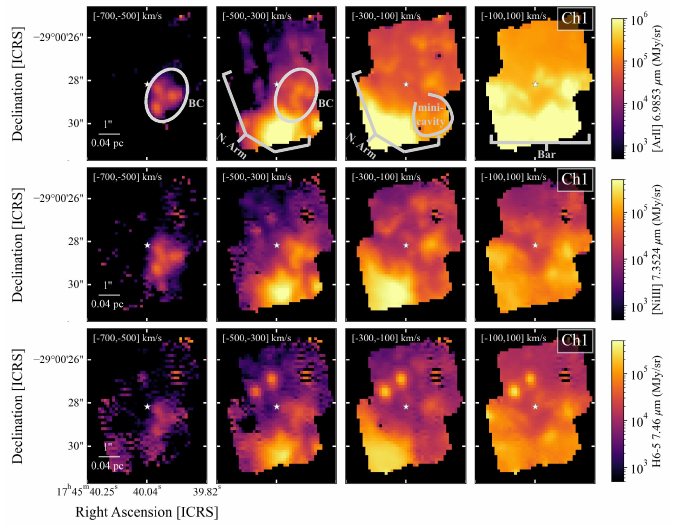}
    \includegraphics[width=0.75\textwidth]{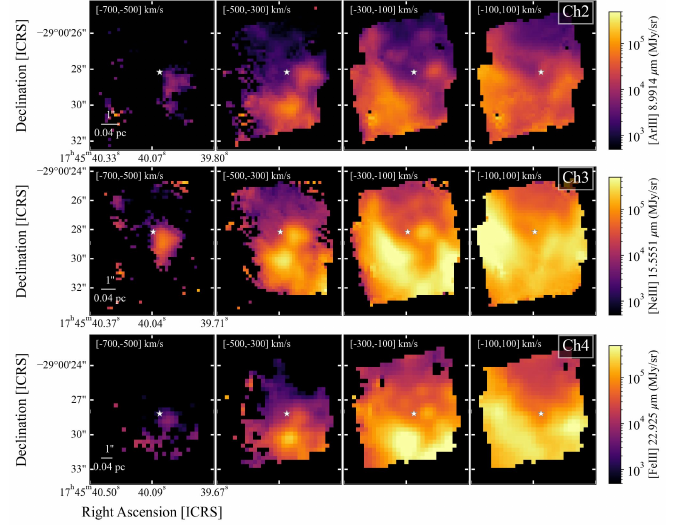}
    \caption{\footnotesize Channel maps at different \vrad\ bins. The white star marks the position of \Sg. Each row shows emission from a single continuum-subtracted line (see colorbar labels), and each panel shows a different \vrad\ bin. Relevant Minispiral structures are labeled in gray in the top row, and additional objects are labeled in Figure~\ref{fig:zoomin}.
    The rightmost column shows emission associated with the Bar, the middle two columns with the Northern Arm (and some of the BC structures), and the leftmost column the BC structures.}
    \label{fig:ChMaps}
\end{figure*}

To generate images that map out superposed emitting structures moving at different \vrad, each emission line's continuum-subtracted, native-resolution cube is spectrally summed over a set of \vrad\ bins (``channels"). The channel-map range (\vrad~$\in[-700, +100]$~km/s) includes the majority of each line's flux, and the 200~\kms\ channel width is comparable to the average fitted line $\sigma \sim 100$--150~\kms. Example channel maps are shown in Figure~\ref{fig:ChMaps} for lines with IPs ranging from 7.46--40.96~eV. 

\textit{Bar $[-100,100]$~\kms}:
The spatial morphology and velocity structure match the properties of the Minispiral Bar \citep{Ekers1983,Paumard2004,Zhao2009}. \cite{Paumard2004} and \cite{Nitschai2020} constrain $\vrad \sim 0$--50~\kms\ for the portion of the Bar in our FoV. The Eastern Arm may also appear at the edges of the FoV (Figure~\ref{fig:zoomin}), but we cannot clearly distinguish it from the Bar in our FoV. 

\textit{Northern Arm $[-300,-100]$~\kms}:
We associate a prominent curved feature with the Northern Arm, which wraps across the FoV from east to west. Although it peaks in the $[-300,-100]$~\kms\ bin, portions of it appear across all \vrad\ bins. Multiple smaller-scale features appear within and around the Northern Arm. The mini-cavity \citep{YusefZadeh1989} is located in the southwestern corner (Figure~\ref{fig:ChMaps} third column). A bright clump is positioned $\sim$1\arcsec\ west of \Sg, overlapping with the known $\epsilon$ source \citep[see  Figure~\ref{fig:zoomin};][]{YusefZadeh1990,Zhao1991}. 

\textit{BC Structures $[-500,-300]$ and $[-700,-500]$~\kms\ Bins}:
Several extended and intersecting (in projection) blueshifted compact (BC) structures appear. We identify X7 based on its position and velocity aligning with orbital model predictions \citep[][tip position marked with a circle in Figure~\ref{fig:zoomin}]{Ciurlo2023}. The X7 tip first appears in the $[-500,-300]$~\kms\ bin, and the tail becomes clear in the $[-700,-500]$~\kms\ bin. Our observations represent the most up-to-date measurement of this object and confirm the \cite{Ciurlo2023} model. Additional BC structures form fast-moving `bridges' connecting the projected positions of \Sg, the tip of X7, and the mini-cavity. 

\textit{Stellar Sources}:
Multiple WR stars also stand out in the channel maps (see Figure~\ref{fig:zoomin} for object labels). The brightest WR stars in our FoV are: IRS 16C and NW, IRS 33N, IRS 21, IRS 29N, and IRS 3E (the latter two objects contain pixels that were masked out in the \jwst\ calibration pipeline). The IRS 13 cluster \citep[e.g.,][]{Maillard2004} appears at the western edge of our FoV (\citealt{Roychowdhury2025} discuss IRS~13E in more detail).

\section{Data Artifacts}\label{sec:artifacts}
\subsection{[Fe III] and [Ni III] Rest Wavelengths}\label{app:FeIII}
Compared to the other emission lines, [\ion{Fe}{3}] shows a redshift offset of $\sim$60~\kms\ across all apertures and gas structures (Figures~\ref{fig:apertures}~and~\ref{fig:Region1fits}). This is likely an artifact of a poorly constrained rest wavelength because a physical Doppler shift should vary spatially and between different gas structures. Based on the redshift, the line's rest wavelength seems closer to 22.930 \um, whereas NIST\footnote{\url{https://physics.nist.gov/PhysRefData/ASD/lines_form.html}} reports 22.925 \um. This redshift does not appear in other Ch4 lines, so it is unlikely to be a result of poor spectral resolution or MIRI wavelength mis-calibration. 

This would not be the first time that a rest wavelength needs to be updated. [\ion{Ni}{3}]'s rest wavelength as given by NIST is 7.349 \um, but \cite{VanDePutte2024} correct it by $\sim$+0.003 \um\ (the revised value, 7.352 \um, is used in this work). The measured [\ion{Fe}{3}] discrepancy of 0.005 \um\ is therefore not unprecedented, but spectral observations of other sources would be helpful to clarify the origins of this apparent discrepancy.

\subsection{[Ar II] Cruciform Artifact}\label{subsec:cruciform}
\begin{figure}
    \centering
\includegraphics[width=0.65\linewidth]{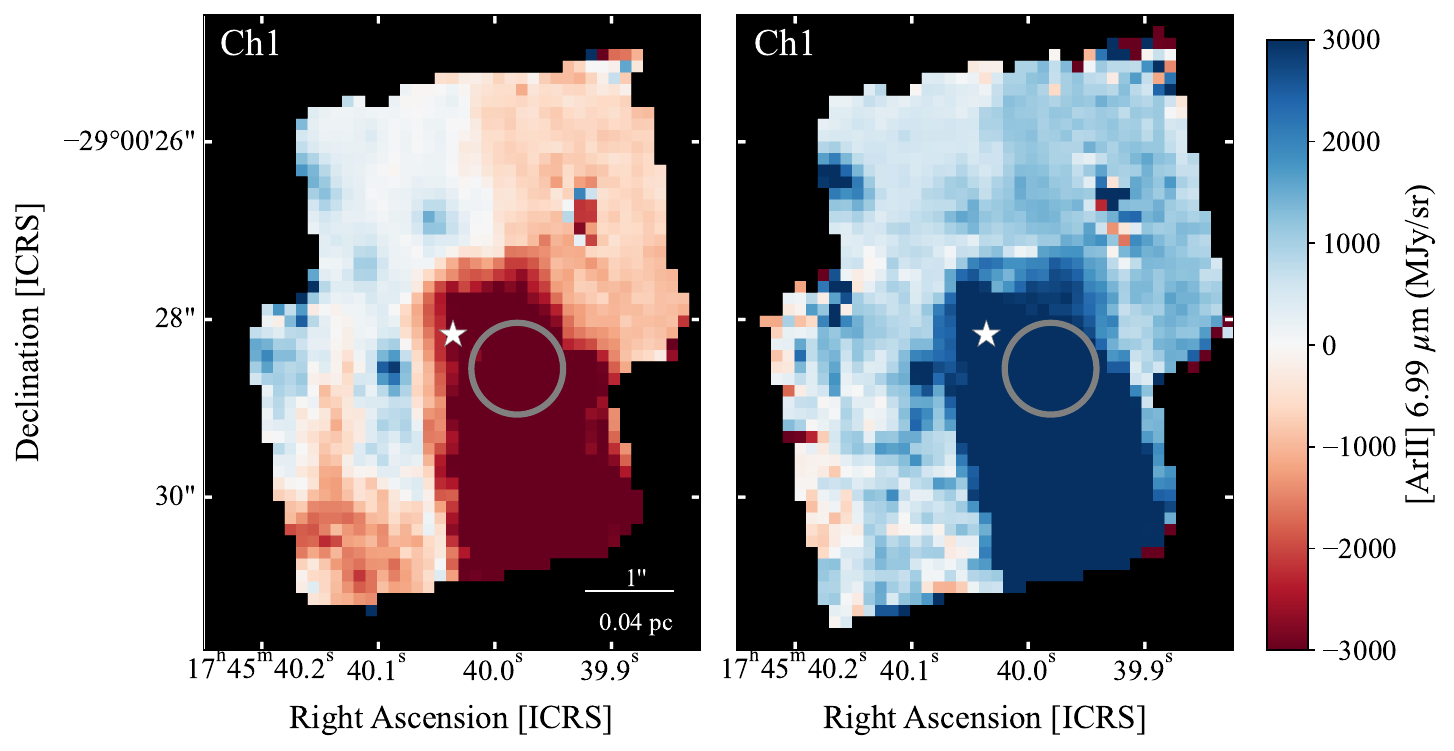}
\includegraphics[width=0.32\linewidth]{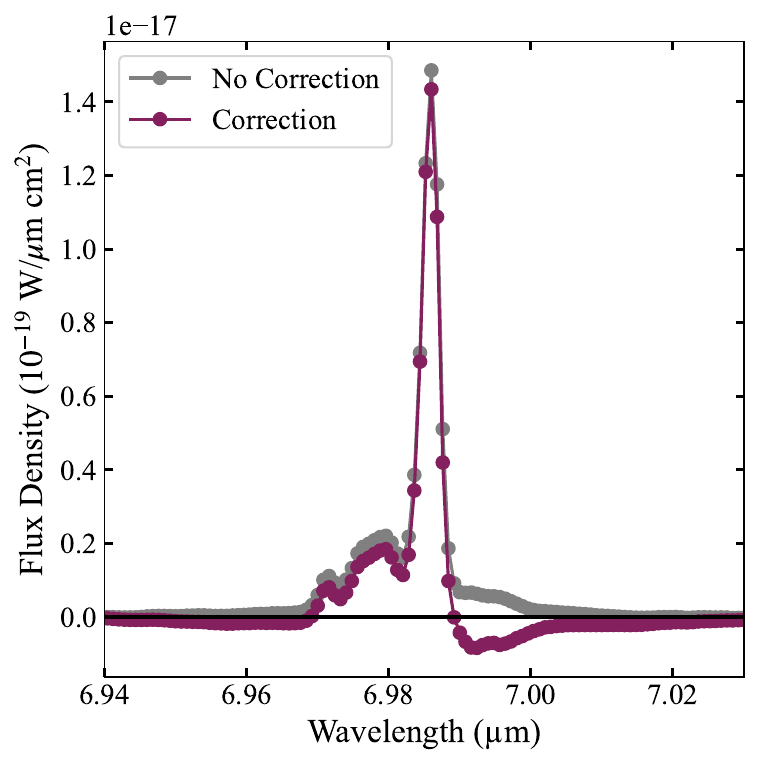}
    \caption{Testing the \jwst\ calibration pipeline \texttt{straylight} correction. The left two panels show a continuum-subtracted, native-resolution spatial slice taken near the [\ion{Ar}{2}] wavelength $\sim$6.99 \um, \textit{left}: with \texttt{straylight} enabled and \textit{middle}: with \texttt{straylight} disabled. Red pixels indicate negative flux (oversubtraction), and blue pixels indicate uncorrected cruciform PSF scattering. \textit{Right}: \texttt{Straylight}-corrected (maroon) and uncorrected (gray) spectra extracted from the aperture marked in the images by the gray circle, positioned over the cruciform artifact `stripe'. The Bar, Northern Arm, and BC structures are minimally impacted by the \texttt{straylight} oversubtraction in our chosen apertures (Figure~\ref{fig:zoomin}).}
    \label{fig:Cruciform}
\end{figure}

The MIRI MRS PSF can scatter bright line features into neighboring slices on the detector. This stray-light signal is automatically removed by the \texttt{straylight} step in the \jwst\ calibration pipeline, but very bright lines in complicated, resolved fields can have a slight over-subtraction that shows up as a dip in flux. Spectrally, this dip manifests most prominently in the bright line's wings, and this issue is identified as the `cruciform artifact' \citep{Argyriou2023}.\footnote{\url{https://jwst-docs.stsci.edu/known-issues/miri-known-issues\#MIRIKnownIssues-cruciformCruciformartifact}} The cruciform artifact is most likely to appear in Ch1 and Ch2; indeed the bright [\ion{Ar}{2}] line extracted from Regions 1 and 2 shows a small flux dip ($\lesssim$5\% of the peak flux) in its red wing (left panel of Figure~\ref{fig:Cruciform}). This absorption dip appears in both our science and the background observation files, and therefore it cannot be corrected by background subtraction.
Other lines are 1--2 dex fainter than [\ion{Ar}{2}]), too faint to be noticeably impacted (Table~\ref{tab:lines_ip}).

To diagnose the impact of the cruciform artifact and its over-subtraction on our line fits, we re-run the calibration pipeline without the \texttt{straylight} step enabled. 
In the corrected version (Figure~\ref{fig:Cruciform} left panel), the over-subtraction spatially appears as negative flux and spectrally appears as a dip in the red wing of the line. Meanwhile, in the uncorrected version (Figure~\ref{fig:Cruciform} middle panel), there is an artificial `stripe' of enhanced flux due to the cruciform artifact's influence on the PSF and a corresponding bump in the spectrum. 

For both the corrected and un-corrected data cubes, we fit a spectrum extracted from an aperture positioned over the cruciform stripe using the procedure outlined in Section~\ref{sec:methods} (Figure~\ref{fig:Cruciform} right panel). The cruciform artifact does not significantly alter the resulting fit parameters, and $X_\mathrm{Ar}$ changes by $\lesssim$5\% for the Bar and Northern Arm. For the BC structures, using the uncorrected data would give 25\% higher $X_\mathrm{Ar}$. 

\subsection{[Ne II] Line Saturation}\label{app:NeII}
The bright [\ion{Ne}{2}] 12.81 \um\ line requires a customized model fit. [\ion{Ne}{2}] is bright enough to saturate the detector in certain portions of the FoV. Regions 1 and 2 avoid saturated pixels, but Region 3 has several saturated pixels arising from the blue-shifted Northern Arm. These saturated pixels are masked out in the cube. 

To account for the saturated wavelengths, we impose stricter fit priors for the Region 3 spectrum. First, we fit the [\ion{Ne}{3}] 15.56 \um\ and [\ion{S}{3}] 18.71 \um\ lines; [\ion{Ne}{3}] because it is the same element as [\ion{Ne}{2}] and [\ion{S}{3}] because its IP is similar to [\ion{Ne}{2}]'s (Table~\ref{tab:lines_ip}). [\ion{Ne}{2}]'s $\mu$ and $\sigma$ bounds are chosen to encompass the corresponding $\mu$ and $\sigma$ obtained from [\ion{Ne}{3}] and [\ion{S}{3}]. The saturated Northern Arm's [\ion{Ne}{2}] amplitude is constrained to be greater than the unsaturated Bar's. We fit [\ion{Ne}{2}] excluding the saturated wavelengths, and we report the flux of the Northern Arm as a lower bound.

\section{Full Line-Fit Results}\label{app:linefits}
\begin{figure*}
    \centering
    \includegraphics[width=\textwidth]{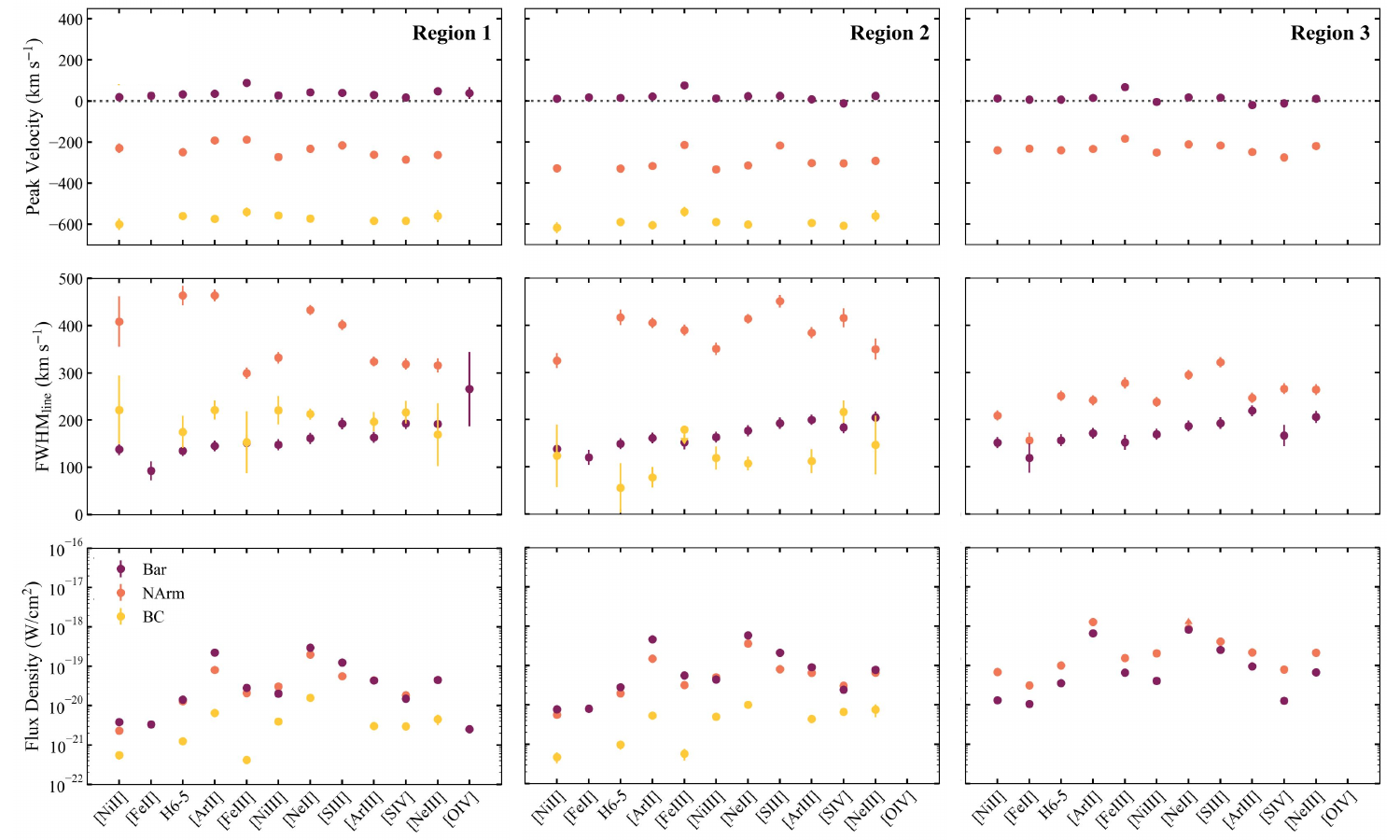}
    \caption{\footnotesize Line-fit results extracted from the Bar (purple), Northern Arm (orange), and the BC structures (yellow) in Regions 1--3. Ionic and recombination lines are ordered by increasing IP (wavelengths and IPs are in Table~\ref{tab:lines_ip}). The top row shows $v_\mathrm{rad}$, the middle row shows line width (FWHM$_{\mathrm{line}}$), and the bottom row shows integrated flux. Results are included for all gas structures with ${>} 5\sigma$ confidence. Error bars are shown, but some are smaller than the point sizes. The Minispiral gas structures are distinguished by different Doppler shifts, while FWHM$_{\mathrm{line}}$ varies substantially for different lines and Regions. Fluxes generally peak for lines with IP near 20~eV.}
    \label{fig:Region1fits}
\end{figure*}

Line fit results sorted by IP are summarized in Figure~\ref{fig:Region1fits} with the numerical form provided in Table~\ref{app:tab:fullfits}.
FWHM$_{\mathrm{line}} \sim 100$--300~\kms\ for all Minispiral structures. For the Bar, there may be a weak trend of increasing FWHM$_{\mathrm{line}}$ with IP (middle row of Figure~\ref{fig:Region1fits}).  However, it is difficult to determine the significance of this trend without FWHM$_{\mathrm{line}}$ measurements for lines with higher IP than [\ion{O}{4}]. For further examples of this trend observed in, e.g., active galactic nuclei outflows, see \cite{Goold2024}. The Northern Arm and BC structures both show significant FWHM$_{\mathrm{line}}$ scatter in Regions 1 and 2. For the Northern Arm, FWHM$_{\mathrm{line}}$ may be slightly inflated due to multiple spectrally unresolved and blended sub-components originating from the Northern Arm's upper branch (Figures~\ref{fig:zoomin} and ~\ref{fig:ChMaps}). For the BC structures, FWHM$_{\mathrm{line}}$ is the least constrained due to the BC structures' faintness but is generally $\lesssim$250~\kms\ with no clear correlation to IP.

\begin{longrotatetable}
\begin{deluxetable}{ccrccccccccc}
\addtolength{\tabcolsep}{-2pt}
\tablecaption{\footnotesize 
Line-fit results for spectra extracted from Regions 1--3. \label{app:tab:fullfits}}
\tablehead{
\colhead{} &
\colhead{} &
\colhead{} &
\multicolumn{3}{c}{\textbf{Region 1}} &
\multicolumn{3}{c}{\textbf{Region 2}} &
\multicolumn{3}{c}{\textbf{Region 3}} \\
\colhead{Transition} &
\colhead{$\lambda_\mathrm{rest}$} &
\colhead{IP} &
\colhead{\vrad} &
\colhead{FWHM$_\mathrm{line}$} &
\colhead{Integrated Flux} &
\colhead{\vrad} &
\colhead{FWHM$_\mathrm{line}$} &
\colhead{Integrated Flux} &
\colhead{\vrad} &
\colhead{FWHM$_\mathrm{line}$} &
\colhead{Integrated Flux}\\
\colhead{} &
\colhead{[\um]} &
\colhead{[eV]} &
\colhead{[km/s]} &
\colhead{[km/s]} &
\colhead{[10$^{-21}$ W cm$^{-2}$]} &
\colhead{[km/s]} &
\colhead{[km/s]} &
\colhead{[10$^{-21}$ W cm$^{-2}$]} &
\colhead{[km/s]} &
\colhead{[km/s]} &
\colhead{[10$^{-21}$ W cm$^{-2}$]}\\
}
\startdata
\lbrack \ion{Fe}{2}\rbrack & $5.340$ & $7.90$ & $29 \pm 14$ & $195 \pm 11$ & $22.01 \pm 3.27$ & $26 \pm 14$ & $163 \pm 11$ & $24.69 \pm 3.66$ & $20 \pm 14$ & $150 \pm 11$ & $32.07 \pm 4.76$ \\
 &  &  & \nodata & \nodata & \nodata & \nodata & \nodata & \nodata & $-241 \pm 14$ & $200 \pm 11$ & $16.28 \pm 2.43$ \\
\lbrack \ion{Ni}{2}\rbrack & $6.636$ & $7.64$ & $19 \pm 14$ & $138 \pm 13$ & $3.80 \pm 0.39$ & $11 \pm 14$ & $138 \pm 12$ & $7.77 \pm 0.75$ & $11 \pm 14$ & $151 \pm 12$ & $13.12 \pm 1.28$ \\
 &  &  & $-230 \pm 22$ & $409 \pm 53$ & $2.31 \pm 0.31$ & $-328 \pm 16$ & $325 \pm 16$ & $5.62 \pm 0.57$ & $-240 \pm 14$ & $209 \pm 11$ & $68.65 \pm 6.58$ \\
 &  &  & $-600 \pm 30$ & $222 \pm 78$ & $0.55 \pm 0.14$ & $-618 \pm 26$ & $123 \pm 69$ & $0.47 \pm 0.15$ & \nodata & \nodata & \nodata \\
\lbrack \ion{Ar}{2}\rbrack & $6.985$ & $15.80$ & $35 \pm 13$ & $145 \pm 11$ & $222.16 \pm 21.97$ & $21 \pm 13$ & $161 \pm 11$ & $468.04 \pm 46.27$ & $15 \pm 13$ & $171 \pm 11$ & $659.01 \pm 65.14$ \\
 &  &  & $-192 \pm 15$ & $464 \pm 13$ & $80.49 \pm 8.04$ & $-317 \pm 14$ & $405 \pm 11$ & $150.48 \pm 14.91$ & $-233 \pm 13$ & $241 \pm 11$ & $1281.43 \pm 126.66$ \\
 &  &  & $-573 \pm 17$ & $221 \pm 20$ & $6.53 \pm 0.77$ & $-606 \pm 16$ & $78 \pm 22$ & $5.39 \pm 0.74$ & \nodata & \nodata & \nodata \\
\lbrack \ion{Ni}{3}\rbrack & $7.352$ & $18.20$ & $27 \pm 14$ & $148 \pm 12$ & $20.38 \pm 1.81$ & $13 \pm 14$ & $163 \pm 12$ & $44.76 \pm 3.96$ & $-5 \pm 14$ & $169 \pm 12$ & $41.06 \pm 3.66$ \\
 &  &  & $-272 \pm 15$ & $332 \pm 12$ & $30.85 \pm 2.75$ & $-334 \pm 15$ & $350 \pm 13$ & $50.80 \pm 4.53$ & $-252 \pm 14$ & $237 \pm 11$ & $206.00 \pm 18.07$ \\
 &  &  & $-558 \pm 20$ & $221 \pm 29$ & $3.97 \pm 0.53$ & $-590 \pm 17$ & $119 \pm 24$ & $4.98 \pm 0.73$ & \nodata & \nodata & \nodata \\
H 6--5 & $7.460$ & $13.60$ & $32 \pm 14$ & $135 \pm 12$ & $14.35 \pm 1.22$ & $15 \pm 14$ & $149 \pm 12$ & $28.32 \pm 2.40$ & $7 \pm 14$ & $156 \pm 12$ & $35.58 \pm 3.08$ \\
 &  &  & $-248 \pm 16$ & $464 \pm 21$ & $12.87 \pm 1.14$ & $-330 \pm 16$ & $417 \pm 16$ & $19.53 \pm 1.74$ & $-240 \pm 14$ & $250 \pm 11$ & $101.15 \pm 8.55$ \\
 &  &  & $-560 \pm 20$ & $175 \pm 34$ & $1.24 \pm 0.24$ & $-590 \pm 22$ & $55 \pm 60$ & $0.95 \pm 0.45$ & \nodata & \nodata & \nodata \\
\lbrack \ion{Ar}{3}\rbrack & $8.991$ & $27.60$ & $30 \pm 14$ & $163 \pm 11$ & $43.71 \pm 4.82$ & $8 \pm 14$ & $200 \pm 11$ & $91.53 \pm 10.11$ & $-21 \pm 14$ & $219 \pm 11$ & $95.08 \pm 10.56$ \\
 &  &  & $-262 \pm 14$ & $324 \pm 11$ & $44.50 \pm 4.91$ & $-304 \pm 14$ & $384 \pm 12$ & $65.18 \pm 7.23$ & $-249 \pm 14$ & $246 \pm 11$ & $217.67 \pm 24.03$ \\
 &  &  & $-583 \pm 16$ & $197 \pm 21$ & $3.01 \pm 0.38$ & $-594 \pm 17$ & $112 \pm 25$ & $4.36 \pm 0.66$ & \nodata & \nodata & \nodata \\
\lbrack \ion{S}{4}\rbrack & $10.511$ & $34.80$ & $18 \pm 14$ & $193 \pm 12$ & $14.98 \pm 1.50$ & $-12 \pm 14$ & $183 \pm 12$ & $24.36 \pm 2.47$ & $-12 \pm 17$ & $166 \pm 22$ & $12.65 \pm 1.57$ \\
 &  &  & $-285 \pm 15$ & $319 \pm 12$ & $18.30 \pm 1.83$ & $-305 \pm 16$ & $416 \pm 20$ & $30.61 \pm 3.19$ & $-276 \pm 14$ & $266 \pm 12$ & $79.36 \pm 7.94$ \\
 &  &  & $-583 \pm 18$ & $217 \pm 24$ & $2.99 \pm 0.37$ & $-608 \pm 18$ & $217 \pm 24$ & $6.60 \pm 0.92$ & \nodata & \nodata & \nodata \\
H 9--7 & $11.309$ & $13.60$ & $19 \pm 17$ & $140 \pm 23$ & $1.42 \pm 0.20$ & $7 \pm 19$ & $174 \pm 30$ & $3.40 \pm 0.51$ & $-6 \pm 26$ & $215 \pm 50$ & $4.81 \pm 1.17$ \\
 &  &  & $-331 \pm 20$ & $461 \pm 34$ & $3.74 \pm 0.45$ & $-359 \pm 22$ & $461 \pm 42$ & $7.97 \pm 0.99$ & $-281 \pm 18$ & $363 \pm 22$ & $24.08 \pm 2.79$ \\
H 7--6 & $12.372$ & $13.60$ & $32 \pm 17$ & $88 \pm 53$ & $2.41 \pm 0.46$ & $20 \pm 14$ & $72 \pm 35$ & $3.99 \pm 0.70$ & $-11 \pm 23$ & $52 \pm 64$ & $2.99 \pm 0.83$ \\
 &  &  & $-232 \pm 21$ & $460 \pm 38$ & $2.73 \pm 0.27$ & $-322 \pm 19$ & $400 \pm 30$ & $4.23 \pm 0.40$ & $-228 \pm 16$ & $207 \pm 17$ & $17.03 \pm 1.47$ \\
\lbrack \ion{Ne}{2}\rbrack & $12.813$ & $21.60$ & $42 \pm 13$ & $162 \pm 12$ & $298.86 \pm 22.16$ & $24 \pm 13$ & $176 \pm 12$ & $591.29 \pm 43.83$ & $18 \pm 14$ & $186 \pm 11$ & $818.32 \pm 61.09$ \\
 &  &  & $-233 \pm 14$ & $433 \pm 10$ & $198.69 \pm 14.74$ & $-314 \pm 14$ & $414 \pm 10$ & $364.83 \pm 27.04$ & $-211 \pm 14$ & $295 \pm 11$ & $>818$ \\
 &  &  & $-572 \pm 14$ & $213 \pm 12$ & $15.72 \pm 1.20$ & $-603 \pm 14$ & $108 \pm 15$ & $10.08 \pm 0.78$ & \nodata & \nodata & \nodata \\
\lbrack \ion{Ne}{3}\rbrack & $15.555$ & $40.96$ & $49 \pm 14$ & $192 \pm 13$ & $45.06 \pm 3.08$ & $24 \pm 14$ & $204 \pm 13$ & $78.17 \pm 5.43$ & $11 \pm 15$ & $206 \pm 13$ & $67.28 \pm 4.71$ \\
 &  &  & $-263 \pm 15$ & $316 \pm 15$ & $45.52 \pm 3.30$ & $-292 \pm 16$ & $350 \pm 22$ & $66.68 \pm 5.42$ & $-220 \pm 14$ & $264 \pm 11$ & $212.21 \pm 14.15$ \\
 &  &  & $-560 \pm 29$ & $169 \pm 67$ & $4.55 \pm 1.36$ & $-561 \pm 28$ & $146 \pm 62$ & $7.55 \pm 2.72$ & \nodata & \nodata & \nodata \\
\lbrack \ion{Fe}{2}\rbrack & $17.936$ & $7.90$ & $26 \pm 15$ & $93 \pm 20$ & $3.34 \pm 0.22$ & $17 \pm 14$ & $120 \pm 16$ & $7.95 \pm 0.49$ & $7 \pm 17$ & $119 \pm 31$ & $10.58 \pm 1.11$ \\
\lbrack \ion{S}{3}\rbrack & $18.713$ & $23.30$ & $40 \pm 14$ & $192 \pm 12$ & $124.61 \pm 6.62$ & $24 \pm 14$ & $192 \pm 12$ & $211.86 \pm 11.31$ & $16 \pm 14$ & $192 \pm 12$ & $251.41 \pm 13.44$ \\
 &  &  & $-215 \pm 14$ & $402 \pm 11$ & $55.96 \pm 2.99$ & $-217 \pm 15$ & $451 \pm 13$ & $81.17 \pm 4.51$ & $-217 \pm 14$ & $322 \pm 11$ & $408.75 \pm 21.74$ \\
H 8--7 & $19.062$ & $13.60$ & $41 \pm 17$ & $190 \pm 23$ & $1.21 \pm 0.14$ & $61 \pm 24$ & $190 \pm 53$ & $2.02 \pm 0.50$ & $104 \pm 23$ & $190 \pm 75$ & $3.32 \pm 0.73$ \\
 &  &  & $-229 \pm 26$ & $450 \pm 53$ & $1.52 \pm 0.17$ & $-191 \pm 34$ & $450 \pm 68$ & $4.05 \pm 0.62$ & $-226 \pm 20$ & $450 \pm 31$ & $13.22 \pm 1.01$ \\
\lbrack \ion{Fe}{3}\rbrack & $22.925$ & $16.19$ & $88 \pm 14$ & $152 \pm 16$ & $28.16 \pm 1.01$ & $75 \pm 14$ & $152 \pm 16$ & $56.17 \pm 2.01$ & $68 \pm 14$ & $152 \pm 16$ & $66.21 \pm 2.39$ \\
 &  &  & $-188 \pm 14$ & $299 \pm 12$ & $21.08 \pm 0.76$ & $-215 \pm 14$ & $390 \pm 12$ & $31.67 \pm 1.15$ & $-184 \pm 14$ & $278 \pm 12$ & $154.54 \pm 5.52$ \\
 &  &  & $-541 \pm 23$ & $153 \pm 66$ & $0.41 \pm 0.06$ & $-541 \pm 24$ & $<179$ & $0.57 \pm 0.19$ & \nodata & \nodata & \nodata \\
 \lbrack \ion{O}{4}\rbrack & $25.890$ & $54.93$ & $39 \pm 28$ & $266 \pm 79$ & $2.53 \pm 0.46$ & \nodata & \nodata & \nodata & \nodata & \nodata & \nodata \\
\lbrack \ion{Fe}{2}\rbrack & $25.988$ & $7.90$ & $26 \pm 16$ & $<235$ & $6.26 \pm 0.80$ & $24 \pm 16$ & $<235$ & $9.90 \pm 1.33$ & $40 \pm 20$ & $<235$ & $13.79 \pm 2.23$ \\
\enddata
\end{deluxetable}
\end{longrotatetable}

\onecolumngrid
\section{Abundance Variations}\label{app:abund}
\subsection{Effect of Different Extinction Curves}\label{app:ext}
\begin{figure}[ht!]
    \centering
    \includegraphics[width=0.75\textwidth]{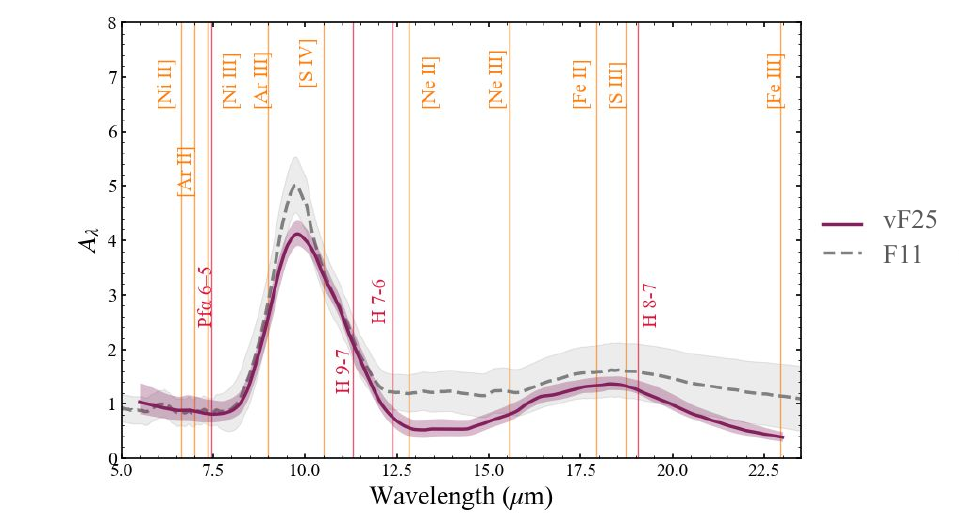}
    \caption{Comparison of the \citetalias{vonFellenberg2025_extinction} (solid purple line) and \citetalias{Fritz2011} (gray dashed line) extinction models.  Shading around each line indicates the 1$\sigma$ confidence intervals. Vertical lines show the relevant hydrogen recombination (red) and ionic lines (orange) used for calculating abundances.}
    \label{fig:extinction}
\end{figure}

The choice of extinction model has a direct impact on the de-reddened line ratios and $X_i$. Though we opt for the \cite{vonFellenberg2025_extinction} model, there are a variety of other GC extinction models  \citep[e.g.,][]{Chiar2006,Fritz2011,Sanders2022}. As a representative example, we compare our findings with the \cite{Fritz2011} model. Both models use the \cite{Kemper2004} dust opacity law. While \citetalias{Fritz2011} fit a 240 K blackbody to ISO/SWS observations of the central 14\arcsec\ $\times$20\arcsec, \citetalias{vonFellenberg2025_extinction} fit the dust continuum model from \cite{Donnan2023,Donnan2024} to the \jwst/MIRI MRS spectrum of the \Sg\ beam. Both models as well as the relevant lines used for calculating the flux ratios and $X_i$ are shown in Figure~\ref{fig:extinction}.

The models share a similar overall shape, differing primarily in the height of the 10 \um\ silicate peak and the flatness of the curve beyond $\sim$12~\um. $A_\lambda$ at all relevant line wavelengths are listed for each model in Table~\ref{tab:extinction}. The calculated abundances depend on the differential extinction ($\Delta A_\lambda$) between a given ionic line and the hydrogen recombination line used for normalization. We select recombination lines that minimize $\Delta A_\lambda$ and improve the measured $X_i$'s robustness to model-dependent $A_\lambda$ variations (Section~\ref{subsec:abundmeth}). For example, [\ion{Ne}{3}] 15.56 \um\ has $A_{15.56,\mathrm{SVF25}} = 0.8$, which differs by almost half a magnitude from $A_{15.56,\mathrm{F11}} = 1.22$. After normalizing to H 7--6 12.37 \um\ with $A_{12.37,\mathrm{SVF25}} = 0.77$ and $A_{12.37,\mathrm{F11}} = 1.22$, however, $\Delta A_{\lambda}$ is near 0 for both models.

If we instead normalize consistently to a single recombination line (e.g., the bright H 6--5 7.46 \um\ line), the \citetalias{vonFellenberg2025_extinction} model predicts smaller $X_\mathrm{Ne}\sim-0.1$ ($0.8\times$ solar), in tension with the inferred super-solar $X_\mathrm{Ar}$. However, when using the same model and normalizing by H 7--6 12.37 \um, $X_\mathrm{Ne} \sim X_\mathrm{Ar} \sim 0.26$ ($1.8\times$ solar). This indicates that the \citetalias{vonFellenberg2025_extinction} model underestimates $\Delta A_{\lambda}$ between H 6--5 and H 7--6 by $\sim$0.7--0.9 mag. Future work is needed to diagnose the cause of this model discrepancy; for the present work, minimizing the differential extinction is sufficient.

\begin{figure*}
    \centering
\includegraphics[width=0.4\linewidth]{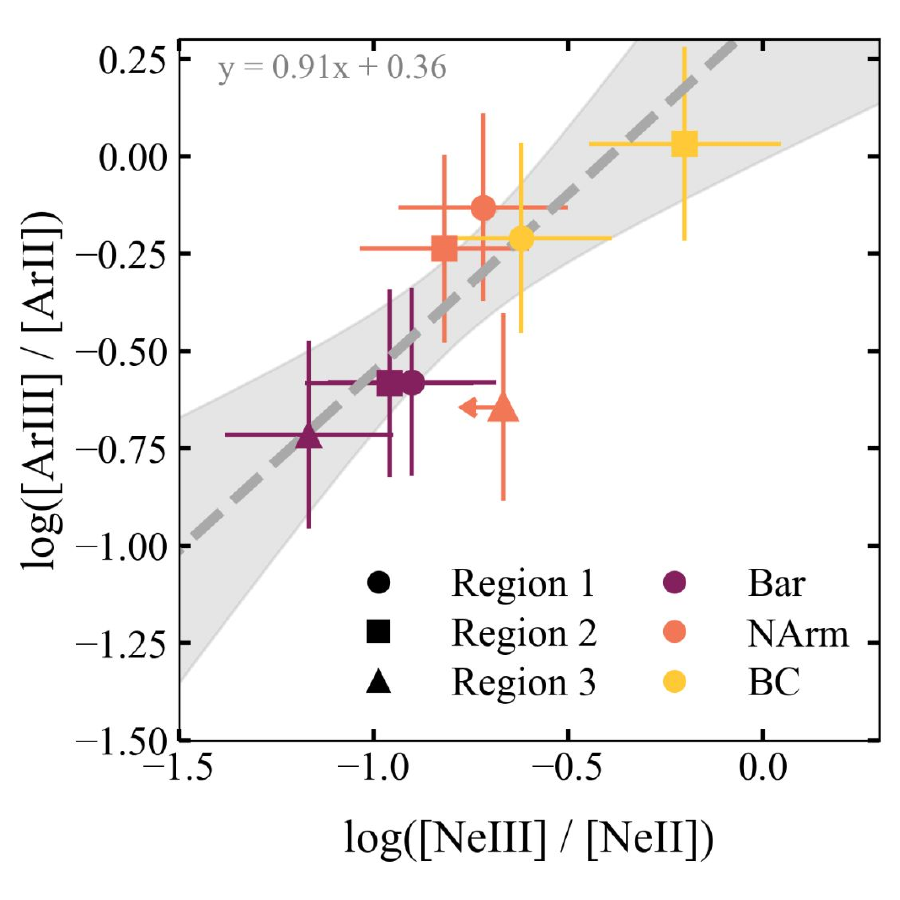}
\includegraphics[width=0.47\linewidth]{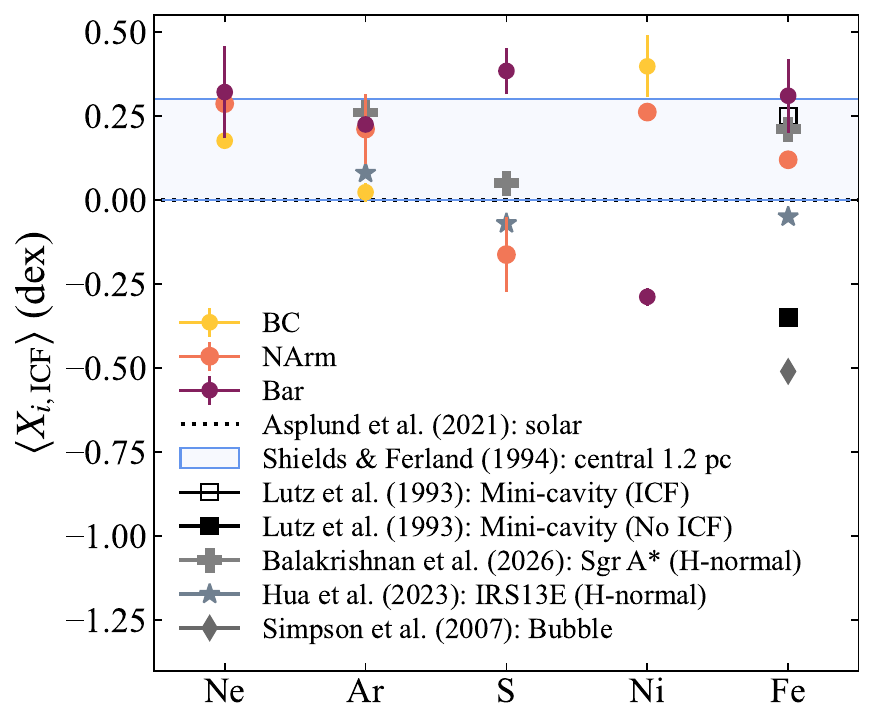}
    \caption{\textit{Left}: line ratio diagram and \textit{Right}: elemental abundances using the \citetalias{Fritz2011}  extinction model instead of the adopted \citetalias{vonFellenberg2025_extinction} model. The left panel corresponds to the upper panel of  Figure~\ref{fig:ratios} and the right panel to  Figure~\ref{fig:AvgAbund}. In both cases, the results are generally consistent using either extinction model.}
    \label{fig:Fritz}
\end{figure*}

The corresponding line ratios and $X_i$ calculated from the \citetalias{Fritz2011} model are displayed in Figure~\ref{fig:Fritz}, presented in the same way as Figures~\ref{fig:ratios}~and~\ref{fig:AvgAbund}. The line ratios (left panel) slope is slightly shallower (0.91 instead of the previous 1.22) but still consistent within the error bars. $X_i$ (right panel) is generally consistent with our findings using \citetalias{vonFellenberg2025_extinction}. The most noticeable change is a slight increase in $\langle X_\mathrm{Ne} \rangle$ and $\langle X_\mathrm{Fe} \rangle$ because of the smaller $\Delta A_{\lambda,\mathrm{F11}}$ compared to $\Delta A_{\lambda,\mathrm{SVF25}}$ for [\ion{Ne}{2}] and [\ion{Fe}{3}] (Table~\ref{tab:extinction}). 

\begin{deluxetable*}{crcccccc}
    \tablecaption{\footnotesize Comparison between the \citetalias{vonFellenberg2025_extinction} and \citetalias{Fritz2011} extinction models \label{tab:extinction}}
    \tablehead{
    \colhead{Transition} & \colhead{$\lambda_{\mathrm{rest}}$} & \colhead{$\mathrm{A}_{\lambda,\mathrm{SVF25}}$} & \colhead{$\mathrm{A}_{\lambda,\mathrm{F11}}$} & \colhead{$|\mathrm{A}_{\lambda,\mathrm{SVF25}}-\mathrm{A}_{\lambda,\mathrm{F11}}|$\tablenotemark{a}} &
    \colhead{Normalization\tablenotemark{b}} & \colhead{$\Delta\mathrm{A}_{\lambda,\mathrm{SVF25}}$\tablenotemark{c}} & \colhead{$\Delta\mathrm{A}_{\lambda,\mathrm{F11}}$\tablenotemark{c}} \\
     & (\um) & (mag) & (mag) & & (mag) & (mag) & (mag) }
     \startdata
     \lbrack \ion{Ni}{2}\rbrack & 6.64 & $0.88^{+0.26}_{-0.18}$ & $0.84\pm0.25$ & 0.04 & H 6--5 & 0.07 & 0.00 \\
     \lbrack \ion{Ar}{2}\rbrack & 6.99 & $0.86^{+0.28}_{-0.18}$ & $0.83\pm0.25$ & 0.03 & H 6--5 & 0.05 & $-0.01$ \\
     \lbrack \ion{Ni}{3}\rbrack & 7.35 & $0.82^{+0.27}_{-0.15}$ & $0.85\pm0.25$ & 0.03 & H 6--5 & 0.01 & 0.01 \\
     H 6--5 & $7.46$ & $0.81^{+0.27}_{-0.15}$ & $0.84\pm0.25$ & 0.03 & \nodata & \nodata & \nodata \\
     \lbrack \ion{Ar}{3}\rbrack & 8.99 & $2.57^{+0.31}_{-0.29}$ & $2.85\pm0.55$ & 0.28 & H 9--7\rlap{\tablenotemark{d}} & 0.46 & 0.66 \\
     \lbrack \ion{S}{4}\rbrack & 10.51 & $3.35^{+0.27}_{-0.21}$ & $3.44\pm0.45$ &  0.09 & H 9--7\rlap{\tablenotemark{d}} & 1.24 & 1.25 \\
     H 9--7 & 11.31 & $2.11^{+0.24}_{-0.24}$ & $2.19\pm0.39$ & 0.08 & \nodata & \nodata & \nodata\\
     H 7--6 & 12.37 & $0.77^{+0.19}_{-0.16}$ &  $1.22\pm0.32$ & 0.45 & \nodata & \nodata & \nodata \\
     \lbrack \ion{Ne}{2}\rbrack & 12.81 & $0.56^{+0.18}_{-0.15}$ & $1.19\pm0.34$ & 0.63 & H 7--6 & $-0.21$ & $-0.03$ \\
     \lbrack \ion{Ne}{3}\rbrack & 15.56 & $0.80^{+0.19}_{-0.11}$ & $1.22\pm0.42$ & 0.42 & H 7--6& 0.03 & 0.00\\
     \lbrack \ion{Fe}{2}\rbrack & 17.94 & $1.33^{+0.16}_{-0.11}$ & $1.58\pm0.48$ & 0.25 & H 8--7 & 0.08 & $-0.01$ \\
     \lbrack \ion{S}{3}\rbrack & 18.71 & $1.33^{+0.16}_{-0.08}$ & $1.60\pm0.50$ & 0.27 & H 8--7 & 0.08 & 0.01\\
     H 8--7 & 19.06 & $1.25^{+0.18}_{-0.07}$ &  $1.59\pm0.51$ & 0.34 & \nodata & \nodata & \nodata \\
     \lbrack \ion{Fe}{3}\rbrack & 22.93 & $0.39^{+0.10}_{-0.08}$ & $1.14\pm0.58$ & 0.75 & H 7--6\rlap{\tablenotemark{d}} & $-0.38$ & $-0.08$ \\
    \enddata
    \vspace{5pt}
    \footnotesize
    \tablenotetext{a}{Difference in $A_{\lambda}$ between the \citetalias{vonFellenberg2025_extinction} and \citetalias{Fritz2011} models (Figure~\ref{fig:extinction}).}
    \tablenotetext{b}{The chosen recombination line that minimizes the differential extinction ($\Delta A_\lambda$) with each ionic line for both extinction models.}
    \tablenotetext{c}{$\Delta A_\lambda$ between the ionic line and the H recombination line chosen for normalization for both extinction models.}
    \tablenotetext{d}{$\Delta A_\lambda$ between the recombination and ionic line is substantial, making the resulting ionic abundance more sensitive to the assumed extinction model.}
\end{deluxetable*}

\onecolumngrid

\subsection{Density and Temperature Dependence}\label{subsec:NeTe}
Increasing \Ne\ beyond $n_\mathrm{crit}$ increases the gas-phase abundance, while increasing $T_e$ decreases the abundance. Low $n_\mathrm{crit}$ ions (Table~\ref{tab:lines_ip}) are more sensitive to \Ne; for this reason, sulfur and iron both increase dramatically in the varied \Ne,\Te\ prescription (Table~\ref{tab:abundancesavg}) because the assumed \Ne\ is larger than in the fixed prescription. 
High $n_\mathrm{crit}$ elements are more impacted by changes in \Te. This is clearest for $\alpha$ elements in the Bar, where the varied prescription's \Te\ is almost double that of the fixed prescription's (Table~\ref{tab:abundancesavg}). 

\begin{figure}[ht]
\centering
\includegraphics[width=\textwidth]{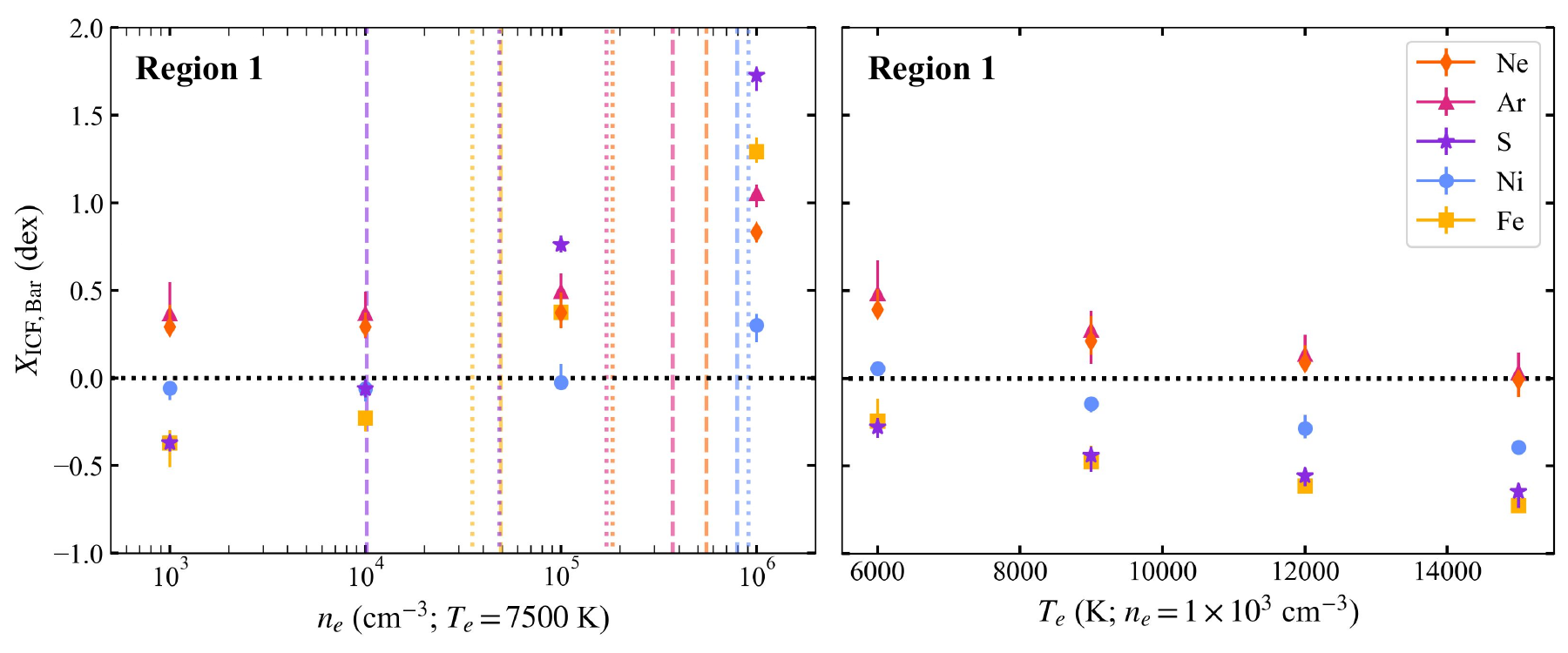}
\caption{\footnotesize Dependence of the derived Region 1 Bar abundances ($X_\mathrm{ICF,Bar}$; log form and solar normalized) on assumed $n_e$ and $T_e$. The left panel shows the effect of varying $n_e = [10^3, 10^6]$~\den\ at fixed $T_e = 7\,500$~K, and the right panel shows varying $T_e = [6\,000, 15\,000]$~K at fixed $n_e = 10^3$~\den. Vertical dashed (dotted) lines in the left panel mark $n_\mathrm{crit}$ for the lower (higher) IP ions used to calculate $X_\mathrm{i,Bar}$ (Table~\ref{tab:lines_ip}). The horizontal dotted line marks solar abundance. Uncertainties are estimated via MC propagation of line flux uncertainties through \texttt{PyNeb}. $X_\mathrm{ICF,Bar}$ increases with \Ne\ and decreases with \Te, and elements with low $n_\mathrm{crit}$ are the most sensitive to the chosen \Ne.}\label{fig:AbundNe}
\end{figure}

To further investigate the abundances' dependence on the assumed \Ne,\Te, we test a series of \Ne,\Te\ combinations that fall within reasonable limits for the GC ionized gas (Section~\ref{subsec:abundmeth}). As an example, we consider the Region 1 Bar abundances with the relevant lower ICF limits applied, hereafter denoted $X_\mathrm{ICF,Bar}$. For one set of calculations, we fix \Te\ $ = 7\,500$ K with varied $n_e = [10^3,10^6]$~\den (left panel of Figure~\ref{fig:AbundNe}). For another, we hold $n_e = 1 \times 10^3$~\den\ with varied $T_e = [6\,000,15\,000]$ K (right panel of Figure~\ref{fig:AbundNe}). For the varied \Ne\ case, iron and sulfur experience the most pronounced change. Argon, neon, and nickel have much higher $n_\mathrm{crit}$ and are minimally impacted except at $n_e \gtrsim 10^6$~\den. Because the identified gas structures all have an assumed $n_e \lesssim 10^5$~\den, those elements should be largely unaffected by \Ne. For the varied \Te\ case, $X_\mathrm{ICF,Bar}$ decreases with increasing \Te. At the highest temperature (15\,000 K), all abundances consistently decrease by a factor of $\sim$1.5--2. Altogether, \Ne\ has a bigger impact on the resulting abundances than \Te\ once \Ne\ surpasses $n_\mathrm{crit}$.

\begin{longrotatetable}
\begin{deluxetable}{@{}c@{\hspace{-15pt}} @{}c@{\hspace{-15pt}} @{}c@{\hspace{-5pt}} cccccccc}
\addtolength{\tabcolsep}{-2pt}
\tablecaption{\footnotesize 
Element abundances for each aperture and gas structure. \label{app1:tab:abundances}}
\tablehead{
\colhead{} &
\colhead{} &
\colhead{} &
\multicolumn{3}{c}{\textbf{Region 1}} &
\multicolumn{3}{c}{\textbf{Region 2}} &
\multicolumn{2}{c}{\textbf{Region 3}} \\
\colhead{Model (\den; K)} &
\colhead{Element} &
\colhead{ICF} &
\colhead{$X_\mathrm{Bar}$} &
\colhead{$X_\mathrm{NArm}$} &
\colhead{$X_\mathrm{BC}$} &
\colhead{$X_\mathrm{Bar}$} &
\colhead{$X_\mathrm{NArm}$} &
\colhead{$X_\mathrm{BC}$} &
\colhead{$X_\mathrm{Bar}$} &
\colhead{$X_\mathrm{NArm}$}
}
    \startdata
    $n_e = 10^3; T_e = 7\,500$ & Ne &
    No & $0.29^{+0.10}_{-0.10}$ & $0.08^{+0.07}_{-0.06}$ & $0.05^{+0.05}_{-0.06}$ & $0.36^{+0.09}_{-0.10}$ & $0.14^{+0.08}_{-0.06}$ & $0.06^{+0.06}_{-0.08}$ & $0.62^{+0.16}_{-0.09}$ & $>-0.10$\\
     & Ar &
     No &
     $0.37^{+0.08}_{-0.10}$ & $-0.03^{+0.08}_{-0.11}$ & $-0.09^{+0.11}_{-0.14}$ & $0.38^{+0.09}_{-0.11}$ & $0.02^{+0.14}_{-0.08}$ & $-0.04^{+0.16}_{-0.13}$ & $0.41^{+0.13}_{-0.11}$ & $0.23^{+0.12}_{-0.11}$\\
     & S &
     No &
     $-0.43^{+0.07}_{-0.05}$ & $-0.87^{+0.06}_{-0.06}$ & \nodata & $-0.43^{+0.13}_{-0.09}$ & $-1.13^{+0.06}_{-0.07}$ & \nodata & $-0.57^{+0.13}_{-0.08}$ & $-0.95^{+0.04}_{-0.06}$\\
      & S & 
      Yes &
      $[-0.37,-0.33]$ & $[-0.81,-0.78]$ & \nodata & $[-0.37,-0.33]$ & $[-1.07,-1.04]$ & \nodata & $[-0.51,-0.48]$ & $[-0.89,-0.86]$\\
     & Ni &
     No &
     $-0.23^{+0.06}_{-0.07}$ & $-0.06^{+0.04}_{-0.06}$ & $0.09^{+0.08}_{-0.08}$ & $-0.19^{+0.08}_{-0.04}$ & $-0.01^{+0.04}_{-0.07}$ & $0.27^{+0.14}_{-0.15}$ & $-0.26^{+0.05}_{-0.05}$ & $-0.01^{+0.05}_{-0.06}$\\
     & Ni & 
     Yes &
     $[-0.06,0.07]$ & $[0.11,0.24]$ & $[0.27,0.39]$ & $[-0.02,0.11]$ & $[0.17,0.29]$ & $[0.45,0.57]$ & $[-0.09,0.04]$ & $[0.16,0.29]$\\
     & Fe &
     No &
     $-0.67^{+0.08}_{-0.06}$ & \nodata & \nodata & $-0.58^{+0.07}_{-0.10}$ & \nodata & \nodata & $-0.42^{+0.12}_{-0.11}$ & $-0.78^{+0.06}_{-0.05}$\\
     & Fe &
     Yes &
     $[-0.37,-0.10]$ & \nodata & \nodata & $[-0.28,-0.01]$ & \nodata & \nodata & $[-0.12,0.14]$ & $[-0.47,-0.21]$\\
     \hline
    $n_{{e,\mathrm{Bar}}} = 10^5; T_{{e,1}} = 13\,000$ & Ne &
     No &
     $0.13^{+0.09}_{-0.10}$ & $0.20^{+0.06}_{-0.07}$ & $0.11^{+0.11}_{-0.10}$ & $0.20^{+0.08}_{-0.06}$ & $0.26^{+0.07}_{-0.05}$ & $0.14^{+0.15}_{-0.10}$ & $0.45^{+0.14}_{-0.11}$ & $>0.02$ \\
    $n_{{e,\mathrm{NArm}}} = 3\times10^4; T_{{e,2}} = 6\,000$ & Ar &
     No &
     $0.22^{+0.11}_{-0.09}$ & $0.12^{+0.12}_{-0.10}$ & $-0.01^{+0.14}_{-0.12}$ & $0.23^{+0.13}_{-0.08}$ & $0.17^{+0.11}_{-0.11}$ & $0.04^{+0.16}_{-0.17}$ & $0.25^{+0.12}_{-0.11}$ & $0.37^{+0.11}_{-0.11}$\\
     $n_{{e,\mathrm{BC}}} = 4\times10^4; T_{{e,3}} = 7\,000$ & S &
     No &
     $0.39^{+0.07}_{-0.06}$ & $-0.08^{+0.06}_{-0.08}$ & \nodata & $0.40^{+0.13}_{-0.08}$ & $-0.34^{+0.08}_{-0.07}$ & \nodata & $0.25^{+0.09}_{-0.10}$ & $-0.16^{+0.06}_{-0.06}$\\
     & S &
     Yes &
     $[0.45,0.49]$ & $[-0.02,0.01]$ & \nodata & $[0.46,0.49]$ & $[-0.28,-0.25]$ & \nodata & $[0.31,0.35]$ & $[-0.10,-0.06]$\\
    & Ni &
     No &
     $-0.47^{+0.05}_{-0.06}$ & $0.05^{+0.06}_{-0.07}$ & $0.13^{+0.07}_{-0.12}$ & $-0.43^{+0.06}_{-0.07}$ & $0.11^{+0.06}_{-0.07}$ & $0.31^{+0.12}_{-0.13}$ & $-0.49^{+0.05}_{-0.06}$ & $0.10^{+0.06}_{-0.06}$\\
     & Ni &
     Yes &
     $[-0.29,-0.16]$ & $[0.23,0.36]$ & $[0.30,0.43]$ & $[-0.25,-0.13]$ & $[0.28,0.41]$ & $[0.49,0.61]$ & $[-0.31,-0.19]$ & $[0.27,0.40]$ \\
    & Fe &
    No &
    $-0.22^{+0.12}_{-0.07}$ & \nodata & \nodata & $-0.13^{+0.08}_{-0.08}$ & \nodata & \nodata & $0.04^{+0.11}_{-0.09}$ & $-0.28^{+0.05}_{-0.06}$\\
    & Fe &
    Yes &
     $[0.08,0.35]$ & \nodata & \nodata & $[0.17,0.44]$ & \nodata & \nodata & $[0.34,0.61]$ & $[0.02,0.28]$\\
     \hline
    \enddata
\tablecomments{Abundances ($X_i$) are expressed in log form relative to solar \citep{Asplund2021}. Section~\ref{subsec:abunds} describes the two different \Ne,\Te\ prescriptions used to calculate $X_i$.}
\end{deluxetable}
\end{longrotatetable}

\onecolumngrid
\subsection{Limitations of MIR Empirical Metallicity Relations}\label{subsec:metal}
Current mid-IR relations for inferring the oxygen abundance ($12+X_\mathrm{O}$ with $X_\mathrm{O}$ in log form and not solar-normalized) may not apply to the extreme conditions of the GC. 
To demonstrate this, we consider the linear relation empirically derived for extragalactic \hii\ regions with sub- to solar-$Z$  \citep{Rogers2026}:
\begin{equation}
   12+X_\mathrm{O} = 1.91 \times \mathrm{Ne}_{23} + 4.87~,
\end{equation}
where Ne$_{23}$ is the log of the sum of the [\ion{Ne}{2}] and [\ion{Ne}{3}] fluxes divided by the H 7--6 flux. 
Plugging in our de-reddened fluxes for the Bar, Northern Arm, and BC structures (Table~\ref{app:tab:fullfits})\footnote{For the BC structures, we use the H 7--6 flux inferred using the method described in Section~\ref{subsec:abundmeth}.} gives $\langle 12+X_\mathrm{O} \rangle \sim 8.7$--9.2 for all the gas structures, i.e., approximately solar \citep[based on][]{Asplund2021}. This is lower than our directly inferred neon abundance (${\sim}1.8\times$solar), suggesting either non-solar $X_\mathrm{Ne}/X_\mathrm{O}$ or systematic errors in applying the Ne$_{23}$ calibration to the extreme environment of the GC. Our assumed \Ne\ and observed Ne$_{23}$ for the Minispiral fall at or beyond the uppermost limits for the calibration sample used by \cite{Rogers2026}. Recognizing that the classical \hii\ scenario cannot fully characterize the ionized gas in the central parsec, we proceed assuming the metal abundances are best represented by the values inferred in Section~\ref{subsec:abunds}. Future work directly measuring the GC gas' oxygen abundance and extending the empirical Ne$_{23}$--$X_\mathrm{O}$ relation to \hii\ regions with higher densities, stronger ionization, and super-solar $Z$ would help clarify how neon and oxygen scale in the GC.

\bibliography{sample7}{}
\bibliographystyle{aasjournalv7}



\end{document}